\documentclass{article}
\usepackage[utf8]{inputenc}
\usepackage{fullpage}
\usepackage{float}
\usepackage{amsmath,amsthm, amsfonts}
\usepackage{graphicx,caption}
\usepackage{xcolor}
\usepackage{authblk}
\usepackage{arydshln}
\usepackage{verbatim}
\usepackage{hyperref}
\usepackage{titlesec}
\usepackage{booktabs}

\theoremstyle{remark}
\newtheorem{remark}{Remark} 
\newtheorem{proposition}{Proposition}

\newcommand{\keywords}[1]{\textbf{\textit{Keywords---}} #1}

\title{Meta-analysis of median survival times with inverse-variance weighting}
\author[1]{Sean McGrath}
\author[1]{Cheng-Han Yang}
\author[2]{Jonathan Kimmelman}
\author[3]{Omer Ozturk}
\author[4]{Russell Steele}
\author[5,6,7]{Andrea~Benedetti}
\date{}

\affil[1]{\small Department of Biostatistics, Yale School of Public Health, New Haven, CT, USA}
\affil[2]{\small Department of Equity, Ethics and Policy, McGill University, Montreal, Quebec, Canada}
\affil[3]{\small Department of Statistics, The Ohio State University, Columbus, OH, USA}
\affil[4]{\small Department of Mathematics and Statistics, McGill University, Montreal, Quebec, Canada}
\affil[5]{\small Department of Epidemiology, Biostatistics, and Occupational Health, McGill University, Montreal, Quebec, Canada}
\affil[6]{\small Respiratory Epidemiology and Clinical Research Unit (RECRU), McGill University Health Centre, Montreal, Quebec, Canada}
\affil[7]{\small Department of Medicine, McGill University, Montreal, Quebec, Canada}

\begin{document}

\maketitle

\begin{abstract}
We consider the problem of meta-analyzing outcome measures based on median survival times. Primary studies with time-to-event outcomes often report estimates of median survival times and confidence intervals based on the Kaplan-Meier estimator. However, outcome measures based on median survival are rarely meta-analyzed, as standard inverse-variance weighted methods require within-study standard errors that are typically not reported. In this article, we consider an inverse-variance weighted approach to meta-analyze median survival times that estimates the within-study standard errors from the reported confidence intervals. We show that this method consistently estimates the standard error of median survival when applied to confidence intervals constructed by the Brookmeyer-Crowley method. We conduct a series of simulation studies evaluating the performance of this approach at the study level (i.e., for estimating the standard error of median survival) and the meta-analytic level (i.e., for estimating the pooled median, difference of medians, and ratio of medians) for commonly used confidence intervals for median survival, including the Brookmeyer-Crowley method and nonparametric bootstrap. We find that this approach often performs comparably to a benchmark approach that uses the true within-study standard errors for meta-analyzing median-based outcome measures when within-study sample sizes are moderately large (e.g., above 50). However, when the effective sample sizes are small, the method can yield biased estimates of within-study standard errors. We illustrate an application of this approach in a meta-analysis evaluating survival benefits of being assigned to experimental arms versus comparator arms in randomized trials for non-small cell lung cancer therapies.
\end{abstract}


\section{Introduction} \label{sec: intro}

Researchers and policy-makers are often tasked with synthesizing time-to-event (survival) data across studies in meta-analyses. For example, decisions about drug reimbursement or clinical practice guidelines may be based on such analyses. When meta-analyses of time-to-event data are performed, the hazard ratio is the typical outcome measure used \cite{higgins2019choosing}. Standard inverse-variance weighted approaches are typically applied to meta-analyze hazard ratios, where study-specific estimates of hazard ratios and their standard errors are often reported or can be estimated from various sets of summary statistics extracted from the primary studies \cite{tierney2007practical}. 

However, the hazard ratio is not the most appropriate outcome measure in some time-to-event studies and meta-analyses of them. Many authors argue that the interpretation of hazard ratios is not straightforward, especially when hazard ratios vary over time (i.e., the assumption of proportional hazards is violated) \cite{uno2014moving, pak2017interpretability, alexander2018hazards, mccaw2022pitfall}. Furthermore, in causal inference settings, (counterfactual) hazard ratios do not have a clear causal interpretation \cite{hernan2010hazards, young2020causal}. In such cases, conclusions from analyses based on hazard ratios can be misleading. Substantive interest may instead lie in quantifying how typical survival times differ between groups, such as the difference of median survival times between groups.

We consider the problem of meta-analyzing outcome measures based on median survival times. Primary studies with time-to-event outcomes often report estimates of median survival times and corresponding confidence intervals due to their interpretability. For example, in a meta-analysis performed by Iskander et al. \cite{iskander2024benefits}, 94\% (120 out of 128) of primary studies reported estimates of median progression-free survival times and confidence intervals separately for cancer patients randomized to experimental arms and comparator arms. In many cases, the full survival curves are not available in the primary studies. For example, this may happen when the outcome of interest in the meta-analysis is a secondary outcome in the primary study. It may also occur when data are extracted from sources such as ClinicalTrials.gov, where survival curves are not reported and full publications may not be available (e.g., \cite{iskander2024benefits}). 

Although outcome measures based on median survival are often of interest and median survival times are routinely reported in time-to-event studies, meta-analyses of median-based outcome measures are relatively uncommon, likely due to methodological challenges that are understudied (see also Section \ref{sec: literature}). For example, the Cochrane Handbook for Systematic Reviews of Interventions \cite{higgins2019choosing} does not discuss methods to meta-analyze outcome measures based on median survival. A key challenge for meta-analyzing an outcome measure based on median survival times is that primary studies rarely report estimates of the standard error of the median survival time estimate, which is needed for applying standard inverse-variance weighted meta-analytic methods. Rather, studies typically report nonparametric confidence intervals around median survival time estimates based on inverting Kaplan-Meier estimates of the survival function \cite{brookmeyer1982confidence}. Unlike confidence intervals for hazard ratios \cite{higgins2019choosing}, confidence intervals for median survival are usually not constructed based on the Wald method. Consequently, the validity of back-computing standard error estimates using the standard Wald formula is not guaranteed. Poor estimation of within-study standard errors can potentially have several negative downstream consequences in meta-analysis, including (i) reducing the efficiency of the pooled outcome estimator, (ii) poor estimation of the between-study heterogeneity in random effects meta-analyses, and (iii) poor confidence interval coverage for the pooled outcome in common effect meta-analyses (e.g., \cite{mcgrath2023standard}). 

In this work, we investigate the validity of an inverse-variance weighted approach to meta-analyzing median survival times that is based on treating confidence intervals around median survival time estimates as Wald-type intervals. We refer to this approach as the \emph{Wald approximation-based approach}. We show that the Wald approximation yields a consistent estimator of the standard error of the median survival estimate for the commonly used Brookmeyer-Crowley confidence interval method \cite{brookmeyer1982confidence}. In simulation studies, we evaluate the finite sample performance of the Wald approximation for estimating standard errors based on different types of confidence intervals. We then investigate the application of the Wald approximation in meta-analysis, where we consider the setting where the meta-analysis consists of one-group studies (i.e., where the target of inference is the median survival time) as well as two-group studies (i.e., where the target of inference is the difference or ratio of median survival times between groups). In additional simulation studies, we  evaluate the performance of the Wald approximation-based approach in meta-analysis and compare it to a benchmark approach that uses the true (unknown) standard errors of the medians. Finally, we illustrate an application of the Wald approximation-based approach in a meta-analysis of median overall survival times of clinical trial participants with non-small cell lung cancer.

\section{Methods}

\subsection{Standard meta-analytic methods} \label{sec: standard methods}

We begin with describing standard aggregate data meta-analytic models and estimators, as the Wald approximation-based approach adopts these assumptions and estimators. We adopt similar notational conventions and terminology as in \cite{mcgrath2024metamedian}.

For each primary study $i$ $(i = 1, \dots, N)$, suppose that the meta-analyst obtains an estimate of the outcome measure $\hat{\theta}_i$ (e.g., the median survival in a single group or the difference of medians between two independent groups) and its standard error $\sigma_i$. The outcome measure is assumed be distributed as
\begin{equation*}
    \hat{\theta}_i \sim \mathrm{Normal}(\theta_i, \sigma_i^2), \quad i = 1, \dots, N,
\end{equation*}
where the $\sigma_i^2$ are considered to be known quantities. In the setting of a common effect meta-analysis, it is assumed that the true study-specific outcome measures, $\theta_i$, are identical, which we denote by $\theta$. In the setting of a random effects meta-analysis, the true study-specific outcome measures are assumed to be distributed as
\begin{equation*}
    \theta_i \sim \mathrm{Normal}(\theta, \tau^2), \quad i = 1, \dots, N.
\end{equation*}
We refer to $\theta$ as the pooled outcome measure and refer to $\tau^2$ as the between-study variance.

The inverse-variance weighted estimator of the pooled outcome measure is given by
\begin{equation*}
    \hat{\theta} = \frac{\sum_{i = 1}^N \hat{\theta}_i w_i}{\sum_{i = 1}^N w_i}, \quad \widehat{\mathrm{SE}}(\hat{\theta}) = \sqrt{\frac{1}{\sum_{i = 1}^N w_i}},
\end{equation*}
where the definition of the weights, $w_i$, depends on whether a common effect or random effects meta-analysis is performed. For common effect meta-analyses, the weights are given by $w_i =  \frac{1}{\sigma_i^2}$. For random effects meta-analyses, the weights are given by $w_i =  \frac{1}{\sigma_i^2 + \hat{\tau}^2}$ where $\hat{\tau}^2$ denotes an estimate of the between-study variance. Estimators of between-study variance have been described and compared elsewhere \cite{viechtbauer2005bias, veroniki2016methods, langan2019comparison}.

\subsection{Wald approximation-based approach}

For study $i$ $(i \in \{1, \dots, N\})$ and group $j$ ($j \in \{1, 2\}$), let $m_{ij}$ denote the true median survival time. For each study $i$ and group $j$, we consider that a Kaplan-Meier estimate of the median survival time (denoted by $\widehat{m}_{ij}$) is extracted along with lower and upper limits of a $100(1-\alpha)\%$ confidence interval (denoted by $l_{ij}$ and $u_{ij}$, respectively). In Section \ref{sec: wald study level}, we describe the Wald approximation-based approach to estimate the standard error of $\widehat{m}_{ij}$ from the extracted data. Then, in Section \ref{sec: wald meta-analytic level}, we describe how this approach can be applied to meta-analyze the median, difference of medians, and ratio of median survival times.

\subsubsection{Study level} \label{sec: wald study level}

\paragraph{Wald intervals}
For study $i$ and group $j$, suppose that the confidence interval for $m_{ij}$ was constructed based on the Wald method (without a transformation). This means that $(l_{ij}, u_{ij})$ were obtained by 
\begin{align}
    l_{ij} & = \widehat{m}_{ij} - z_{1-\alpha/2}\widehat{\mathrm{SE}}(\widehat{m}_{ij}) \label{eq:ci lb} \\
    u_{ij} & = \widehat{m}_{ij} + z_{1-\alpha/2}\widehat{\mathrm{SE}}(\widehat{m}_{ij}) \label{eq:ci ub}
\end{align}
where $z_{1-\alpha/2}$ denotes the $1 - \alpha/2$ quantile of the standard normal distribution and $\widehat{\mathrm{SE}}(\widehat{m}_{ij})$ denotes an estimate of the standard error of $\widehat{m}_{ij}$. For example, $\widehat{\mathrm{SE}}(\widehat{m}_{ij})$ may be the conventional estimator given by $\widehat{\mathrm{SE}}(\widehat{m}_{ij}) = \sqrt{\frac{\hat{\sigma}_{ij}^2(\widehat{m}_{ij})}{n_{ij}\hat{f}_{ij}^2(\widehat{m}_{ij})}}$ 
where $\hat{\sigma}_{ij}^2(\cdot)$ is based on the Greenwood formula, $n_{ij}$ is the sample size, and  $\hat{f}_{ij}(\cdot)$ is an estimator of the probability density function for the event times \cite{andersen1993statistical} (see Section 1 of the Supplementary Material for more details). Following (\ref{eq:ci lb}) and (\ref{eq:ci ub}), we can back-compute $\widehat{\mathrm{SE}}(\widehat{m}_{ij})$ by 
\begin{equation}
    \widehat{\mathrm{SE}}(\widehat{m}_{ij}) = \frac{u_{ij} - l_{ij}}{2z_{1-\alpha/2}} \label{eq: wald two limits}
\end{equation}
which we refer to as the Wald approximation. 

\begin{remark}
    For simplicity, we motivated the Wald approximation based on the setting of using the Wald method without a transformation and with equal tail probabilities (i.e., $P(l_{ij} > m_{ij}) = P(u_{ij} < m_{ij}) = \alpha/2$). It is straightforward to see via Taylor expansions that (\ref{eq: wald two limits}) can also be motivated in scenarios where the confidence interval was constructed on a different scale (e.g., log scale) and was back-transformed to the original scale. Moreover, when the confidence interval has unequal tail probabilities, we illustrate in Section 2 of the Supplementary Material that the Wald approximation will overestimate the standard error although the degree of bias is small unless one of the tail probabilities is nearly 0. 
\end{remark}

\begin{remark}
    An upper limit for the confidence interval for $m_{ij}$ may not be reported. In this case, if the confidence interval was constructed using the Wald method, 
    one can estimate the standard error by $\widehat{\mathrm{SE}}(\widehat{m}_{ij}) = \frac{m_{ij} - l_{ij}}{z_{1-\alpha/2}}$.
\end{remark}

\paragraph{Brookmeyer-Crowley intervals}

As discussed in Section \ref{sec: intro}, confidence intervals for median survival reported in practice are typically not Wald intervals. Rather, such confidence intervals are commonly constructed using the test-inversion method proposed by Brookmeyer and Crowley \cite{brookmeyer1982confidence}. Unlike Wald intervals, Brookmeyer-Crowley intervals are not symmetric around the median estimate, which raises questions about whether the Wald approximation in (\ref{eq: wald two limits}) remains valid. Here, we show that the Wald approximation yields a consistent estimator of the standard error of the Kaplan-Meier estimator of median survival under this confidence interval method.

We first introduce the following notation and assumptions. For simplicity, we drop the study ($i$) and group ($j$) subscripts in this subsection. Let $T_1, \dots, T_n$ denote i.i.d.\ event times with distribution function $F$, density $f$, and survival function $S(t) = 1 - F(t)$. Let $C_1, \dots, C_n$ denote i.i.d.\ censoring times that are independent of the event times. The observed data consist of $Y_k = \min(T_k, C_k)$ and $\delta_k = I(T_k \le C_k)$ for $k = 1, \ldots, n$. The population median survival time is given by $m = S^{-1}(1/2)$. We assume that $f$ is continuous and positive at $m$ (so $S'(m) = -f(m) < 0$), and that $P(Y_k \ge \nu) > 0$ for some $\nu > m$ to ensure identifiability and validity of the asymptotic theory.

Let $\widehat{S}_n(t)$ denote the Kaplan-Meier estimator of the survival function, and let $\widehat{m}_n = \widehat{S}_n^{-1}(1/2)$ denote the Kaplan-Meier estimator of the median. Let $\sigma^2(t)$ denote the asymptotic variance of $\sqrt{n}(\widehat{S}_n(t) - S(t))$, and let $\hat{\sigma}_n^2(t)$ denote its estimator based on the Greenwood formula (see Section 1 of the Supplementary Material for details). The Brookmeyer-Crowley method constructs a $100(1-\alpha)\%$ confidence interval $[L_n, U_n]$ for the median by inverting a hypothesis test. Specifically, the confidence interval is defined as
\[
\left\{ t : \frac{|\widehat{S}_n(t) - 1/2|}{\widehat{\sigma}_n(t)/\sqrt{n}} \le z_{1-\alpha/2} \right\},
\]
with $L_n$ and $U_n$ denoting its infimum and supremum, respectively.
We use uppercase $L_n, U_n$ to denote the confidence interval limits as random variables (emphasizing their dependence on sample size $n$), in contrast to the lowercase $l_{ij}, u_{ij}$ used elsewhere to denote observed confidence limits extracted from studies. This interval may be constructed on the original scale (as shown above) or on a transformed scale (e.g., using log or log-minus-log transformations of the survival function) and then back-transformed.

Under these conditions, the Kaplan-Meier estimator of the median is asymptotically normally distributed:
\begin{equation*}
\sqrt{n}(\widehat{m}_n - m) \xrightarrow{d} \mathrm{Normal}(0, V(m))
\end{equation*}
where $\xrightarrow{d}$ denotes convergence in distribution as $n \to \infty$ and $V(m) = \sigma^2(m)/f^2(m)$. The following result establishes that the Wald approximation (scaled by $\sqrt{n}$) provides a consistent estimator of the asymptotic standard error, $\sqrt{V(m)}$.

\begin{proposition} \label{prop:consistency}
Under the given assumptions, the Wald approximation for the standard error satisfies:
\[
\sqrt{n} \frac{U_n - L_n}{2 z_{1-\alpha/2}} \xrightarrow{p} \sqrt{V(m)}
\]
where $\xrightarrow{p}$ denotes convergence in probability as $n \to \infty$.
\end{proposition}

The proof is provided in Section 3 of the Supplementary Material. The proof uses the stochastic equicontinuity of the Kaplan-Meier process and Taylor expansions of the survival function around $m$. The key insight is that when expressing the interval width $U_n - L_n$, the empirical process terms $\widehat{S}_n(m) - 1/2$ appearing in $U_n$ and $L_n$ cancel out, leaving only a deterministic term involving $V(m)$ and asymptotically negligible remainders. A similar result was shown by Ozturk and Balakrishnan \cite{ozturk2020meta} in non-survival settings (i.e., without censoring) based on Bahadur-type representations of sample quantiles. 

Despite this alternative interpretation, we will refer to the estimator in (\ref{eq: wald two limits}) as the Wald approximation-based approach for simplicity.

\subsubsection{Meta-analytic level} \label{sec: wald meta-analytic level}

We consider applying the standard inverse-variance weighted approach as described in Section \ref{sec: standard methods} to meta-analyze the following outcome measures: the median survival time in a single group, the difference of median survival times between two independent groups, and the ratio of median survival times between two independent groups. 

To meta-analyze any of the outcome measures, we first estimate the standard error of the median estimator (in each group) as described in the previous subsection. One can immediately proceed to meta-analyze the median in a single group. To meta-analyze the difference of medians, we need to estimate the standard error of the difference of medians estimator. Due to the independence of the two groups, we can estimate the standard error by
\begin{equation*}
    \widehat{\mathrm{SE}}(\widehat{m}_{i1} - \widehat{m}_{i2} ) = \sqrt{\widehat{\mathrm{SE}}(\widehat{m}_{i1})^2 + \widehat{\mathrm{SE}}(\widehat{m}_{i2})^2}.
\end{equation*}

To meta-analyze the ratio of medians, we apply a log transformation to the outcome measure at the study level, as is standard for meta-analyzing outcome measures that are ratios (e.g., odds ratios, risk ratios, ratios of means) \cite{higgins2019choosing, friedrich2008ratio}. That is, we use the outcome measure of $\hat{\theta}_i = \log( \widehat{m}_{i1} / \widehat{m}_{i2})$. A key motivation for performing the log transformation is that it helps make the assumption that the outcome measure is normally distributed more tenable. Using the delta method, we can construct an estimate of the standard error of $\log( \widehat{m}_{i1} / \widehat{m}_{i2})$ as follows:
\begin{equation*}
    \widehat{\mathrm{SE}}\left(\log\frac{\widehat{m}_{i1}}{\widehat{m}_{i2}}\right) = \sqrt{\frac{\widehat{\mathrm{SE}}(\widehat{m}_{i1})^2}{\widehat{m}_{i1}^2} + \frac{\widehat{\mathrm{SE}}(\widehat{m}_{i2})^2}{\widehat{m}_{i2}^2}}.
\end{equation*}
After obtaining an estimate of the pooled log ratio of medians, the inverse transformation is applied. That is, the estimate of the pooled ratio of medians is $\exp(\hat{\theta})$. We elaborate on the interpretation of the pooled ratio of medians in Remark \ref{rem: ratio interpretation}.

\begin{remark} \label{rem: ratio interpretation}
    As with any nonlinear transformation, the application of the log transformation to the outcome has implications in the interpretation of the target parameter in the meta-analysis. The pooled ratio of medians, $\exp(\theta)$, can be interpreted as the \emph{median} of the distribution of the study-specific ratios of medians. This quantity generally differs from the \emph{mean} of the study-specific ratios of medians due to the assumptions made on additive between-study heterogeneity on the log scale. Note that this distinction is not relevant for the pooled median or pooled difference of medians under our model assumptions: the pooled (difference of) medians can be interpreted as either the mean or median of the distribution of the study-specific (difference of) medians because they coincide.
\end{remark}

\section{Simulations} \label{sec: simulations}

In this section, we present simulation studies evaluating the performance of the Wald approximation-based approach at both the study level and meta-analytic level. Specifically, we evaluate the performance of the Wald approximation-based approach for estimating the standard error of median survival estimates in Section \ref{sec: study level}. We evaluate the Wald approximation-based approach for meta-analyzing the median, difference of medians, and ratio of median survival times in Section \ref{sec: meta-analytic level}. The code for reproducing these simulation studies is available at  \url{https://github.com/stmcg/ma-median-survival}.

\subsection{Study-level simulations} \label{sec: study level}

\subsubsection{Data generation} \label{sec: study level data}

In these simulations, we considered a total of $3 \times 2 \times 4 = 24$ scenarios by varying the distribution of the underlying event times (3 levels), the distribution of censoring times (2 levels), and the sample size (4 levels). In each scenario, we generated 1000 longitudinal data sets with right censoring.

To generate each longitudinal data set, we simulated $n$ underlying event times ($n \in \{50, 100, 250, 1000\}$) from either (i) an exponential distribution with a mean of 40, (ii) a Weibull distribution with a shape parameter of $k=2$ and a scale parameter of $\lambda=35$, or (iii) a two-component mixture of Weibull distributions. The first component had a probability of $2/3$ with parameters $(k = 2, \lambda=20)$  and the second component had a probability of $1/3$ with parameters $(k = 3/2, \lambda = 50)$. Plots of the density functions of these distributions are given in Section 4.1 of the Supplementary Materials. Individuals were considered administratively lost to follow-up if their underlying event times exceeded 100. The censoring times were independently simulated from either a $\text{Uniform}(0, 100)$ distribution or an exponential distribution with a mean of 60. We observe the minimum of the underlying event time and the censoring time. The percentage of individuals censored for each combination of event time distribution and censoring time distribution is listed in Section 4.1 of the Supplementary Materials. The censoring rates ranged between 26\% and 41\%.

In each longitudinal data set, we estimated the median survival based on the Kaplan-Meier estimator \cite{kaplan1958nonparametric}. We considered that three types of confidence intervals may be reported. The first two are based on the Brookmeyer-Crowley method \cite{brookmeyer1982confidence}, as implemented in the \textbf{survival} R package \cite{survival-package}. The first type is based on using the log transformation of the survival curve and the second is based on using the log-minus-log transformation. The third type of confidence interval is the percentile-based nonparametric bootstrap method \cite{efron1994introduction} with 1000 bootstrap replicates.  

We applied the Wald approximation to estimate the standard error of the median survival estimate for each type of confidence interval in each longitudinal data set. We then compared the estimated standard errors to the true standard errors of the median estimator. Specifically, we computed the relative bias of the standard error estimator. Letting $\widehat{\mathrm{SE}}_i(\widehat{m})$ denote the estimated standard error of the median estimator ($\widehat{m}$) in the $i$th iteration $(i = 1, \dots, 1000)$ and $\mathrm{SE}(\widehat{m})$ denote the true value of the standard error, the relative bias is defined as the average of $(\widehat{\mathrm{SE}}_i(\widehat{m}) - \mathrm{SE}(\widehat{m})) / \mathrm{SE}(\widehat{m}) \times 100\%$ across the $1000$ iterations. The true standard errors were obtained by performing Monte Carlo integration with $10^5$ samples (see Section 4.6 of the Supplementary Material for more details). 

\subsubsection{Results} \label{sec: study level results}

\begin{figure} [H]
    \centering
\includegraphics[width=0.95\textwidth]{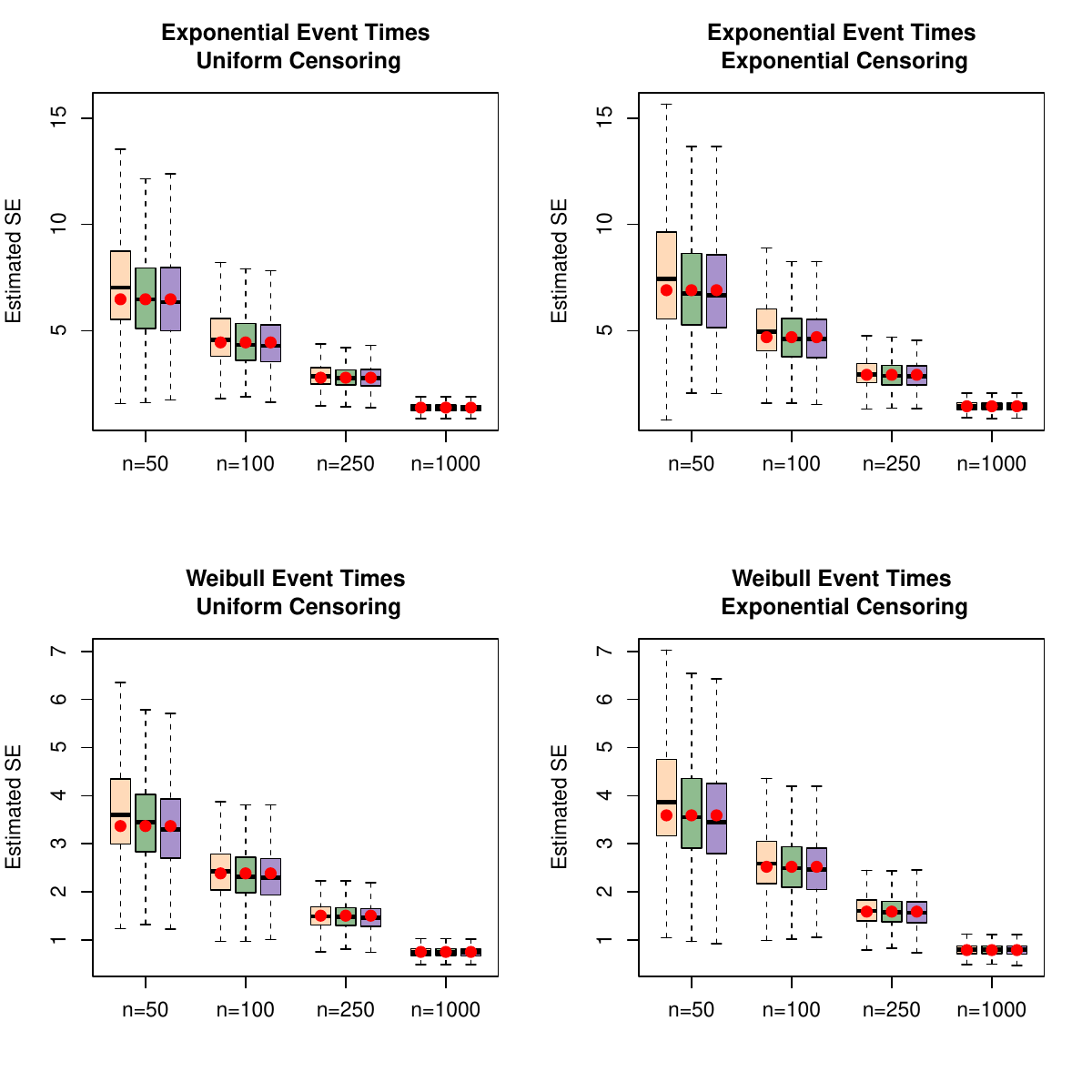}
    \caption{Study-level simulation results for the scenarios with the exponential and Weibull event time distributions. The box plots illustrate the estimated standard errors (SEs) of the median survival time from the Wald approximation-based approach. The peach boxes correspond to when the Brookmeyer-Crowley method based on a log transformation was used to construct the 95\% confidence interval; The green boxes correspond to the Brookmeyer-Crowley method based on a log-minus-log transformation; The purple boxes correspond to the nonparametric bootstrap method. The true standard errors are illustrated by red dots\label{fig:sim res study-level}}
\end{figure}

Figure \ref{fig:sim res study-level} illustrates the distributions of the estimated standard errors in the scenarios with the exponential and Weibull distributions. The corresponding figure for the mixture distribution is placed in Section 4.1 of the Supplementary Material, as similar trends held. Section 4.1 of the Supplementary Material also lists the relative bias values in all scenarios. 

The Wald approximation produced standard error estimates that were often nearly unbiased. The relative bias of this estimator was below 10\% in most of the simulation scenarios. The sample size was the strongest factor affecting the (relative) bias and, as expected, variability of the standard error estimators. In all scenarios, the (relative) bias and variability of the standard error estimator decreased as the sample size increased. When the sample size was 1000, the relative bias of the estimator was below 2\% in all scenarios. 

The standard error estimator was not strongly affected by the choice of event time distribution or censoring time distribution. Although the same trends held across the three different types of confidence intervals used, the standard error estimator had the largest bias when using the Brookmeyer-Crowley confidence interval based on the log transformation.

\subsubsection{Additional study-level simulations} \label{sec: study level adhoc}

We performed a series of additional study-level simulation studies to explore the sensitivity of the Wald approximation to (i) small sample sizes, (ii) a higher degree of censoring, (iii) extremely skewed event time distributions, and (iv) the number of bootstrap replicates when bootstrap confidence intervals are reported.

\paragraph{Impact of smaller sample sizes}

To evaluate the performance of the Wald approximation in settings where the asymptotic normality of the Kaplan-Meier estimator may not serve as a good approximation, we performed additional study-level simulations with smaller sample sizes. Specifically, we considered sample sizes of $n \in \{20, 30, 40\}$ while keeping the same event time and censoring distributions as described in Section \ref{sec: study level data}. 

\begin{figure} [H]
    \centering
\includegraphics[width=0.95\textwidth]{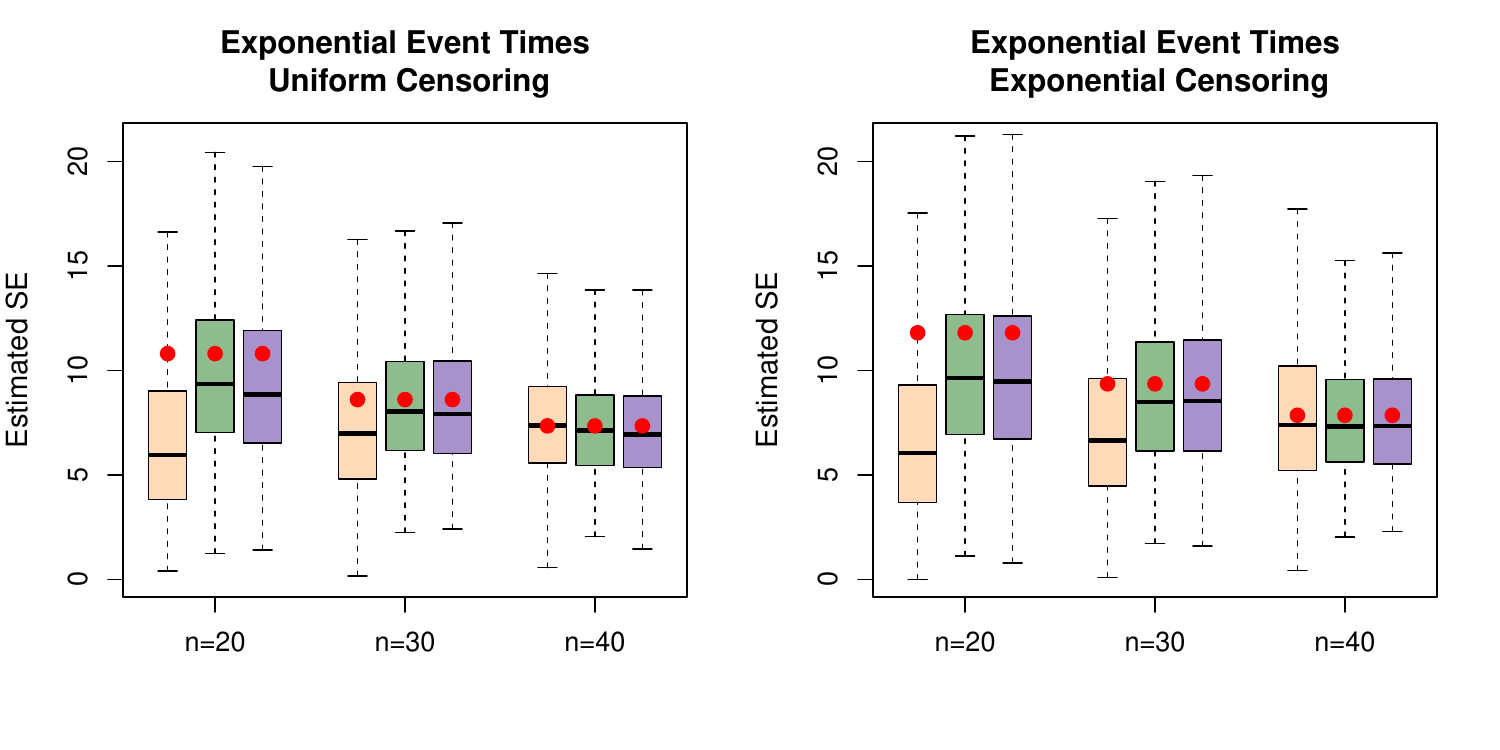}
    \caption{Study-level simulation results for the scenarios with small sample sizes and the exponential event time distribution. The box plots illustrate the estimated standard errors (SEs) of the median survival time from the Wald approximation-based approach. The peach boxes correspond to when the Brookmeyer-Crowley method based on a log transformation was used to construct the 95\% confidence interval; The green boxes correspond to the Brookmeyer-Crowley method based on a log-minus-log transformation; The purple boxes correspond to the nonparametric bootstrap method. The true standard errors are illustrated by red dots\label{fig:sim res study-level smalln}}
\end{figure}

The results for the scenarios with the exponential event time distribution are given in Figure \ref{fig:sim res study-level smalln}, and the corresponding figure for the Weibull distribution and mixture distributions is placed in Section 4.2 of the Supplementary Material. The Wald approximation often exhibited considerable bias when $n = 20$. For example, the relative bias was up to -39\% in one scenario (exponential event time distribution, exponential censoring distribution). In the scenarios with $n = 40$, the bias was significantly reduced, yielding results comparable to the scenarios where $n = 50$. 

\paragraph{Impact of a higher degree of censoring}

We also investigated the impact of having a higher degree of censoring on the performance of the Wald approximation. We modified the data generation mechanism described in Section \ref{sec: study level data} to include higher degrees of censoring by generating censoring times from exponential distributions with means of 40 and 25. We considered the same event time distributions and sample sizes as in Section \ref{sec: study level data}. These scenarios had censoring rates ranging between 43\% and 65\%. 

The results for these scenarios are placed in Section 4.3 of the Supplementary Materials for parsimony. The Wald approximation performed poorly in scenarios with the highest degree of censoring and smallest sample sizes. For example, the relative bias was -39\% in the scenario with high censoring and $n = 50$. However, even in the settings with the highest censoring, the Wald approximation performed well when the sample size was large. For example, the relative bias was below 2\% in all scenarios with $n = 1000$.

\paragraph{Impact of highly skewed event time distributions} We additionally performed simulations with very highly skewed event time distributions in Section 4.4 of the Supplementary Material. When the sample size was small, the Wald approximation clearly showed bias, especially when using the Brookmeyer-Crowley method based on the log transformation. In the most extreme case, a relative bias of 19\% was observed. However, the bias generally decreased as the sample size increased. When the sample size was 1000, the relative bias was below 2\% in all these scenarios as well.

\paragraph{Impact of the number of bootstrap replicates} We investigated sensitivity of the Wald approximation to the number of bootstrap replicates when nonparametric bootstrap confidence intervals are reported. The results are presented in Section 4.5 of the Supplementary Material. Increasing the number of bootstrap replicates did not meaningfully affect the bias or precision of the Wald approximation.

\subsection{Meta-analytic level simulations} \label{sec: meta-analytic level}

We conducted three simulation studies to evaluate the performance of the Wald approximation-based method to meta-analyze the median survival time, difference of median survival times, and ratio of median survival times. These simulation studies are presented in Sections \ref{sec: sim median}, \ref{sec: sim median difference}, and \ref{sec: sim median ratio}, respectively.  

In each simulation study, we considered three scenarios by varying the degree of between-study heterogeneity. In each scenario, we generated 1000 meta-analytic data sets consisting of 20 primary studies. The details of how the primary study data were simulated in each simulation study are given in the respective section. After simulating the primary study data, each study reported an estimate of the median survival based on the Kaplan-Meier method \cite{kaplan1958nonparametric} (for each group, if applicable). We considered that studies may report one of the three types of 95\% confidence intervals as described in the study-level simulations (i.e., two based on the Brookmeyer-Crowley method \cite{brookmeyer1982confidence} with different transformations as well as the percentile-based nonparametric bootstrap method \cite{efron1994introduction}). Each study was randomly assigned with equal probability to report one of these three types of confidence intervals.

We compared the performance of the Wald approximation-based method to a benchmark approach in each simulation study. The benchmark approach applies the inverse-variance method with the true standard errors of the outcome measure, which were obtained by Monte Carlo integration with 1000 samples. We included the benchmark approach to isolate the effect of biased estimation of the within-study standard errors in the Wald approximation-based method. The benchmark method may still have some suboptimal properties (e.g., bias and non-nominal coverage of the confidence intervals) due to some minor violations of the meta-analytic model assumptions. For example, the Kaplan-Meier estimator of the median survival is biased in finite samples. 

In the simulation scenarios with between-study heterogeneity, we used the restricted maximum likelihood (REML) method to estimate the between-study variance \cite{viechtbauer2005bias}. We used the Hartung-Knapp method \cite{hartung2001refined, sidik2002simple} to construct a 95\% confidence interval for the pooled outcome measure and used the Q-profile method \cite{viechtbauer2007confidence} to construct a 95\% confidence interval for the between-study variance as implemented in the \textbf{metafor} package \cite{viechtbauer2010conducting}. In the common effect simulation scenarios, we constructed a confidence interval for the pooled outcome measure by the t-interval $\hat{\theta} \pm t_{N-1, 0.975}\widehat{\mathrm{SE}}(\hat{\theta})$. We evaluated the bias, standard error, and coverage of the 95\% confidence interval for the pooled outcome measure and the between-study variance (if applicable) of the Wald approximation-based approach and benchmark approach. 

\subsubsection{Meta-analysis of median survival} \label{sec: sim median}

\paragraph{Data generation}

We generated longitudinal data of one-group primary studies in a similar manner as in the group/study-level simulations (Section \ref{sec: study level}) with some modifications to allow for between-study heterogeneity. For the $i$th primary study, we generated the sample size from a discrete $\mathrm{Uniform}(50, 1000)$ distribution and generated the underlying event times from an $\mathrm{Exponential}(\lambda_i)$ distribution. The rate parameter $\lambda_i$ was generated so that the true study-specific median survival times follow a normal distribution with mean $\frac{\log 2}{0.025} \approx 27.73$ and variance $\tau^2$. Specifically, we generated $\lambda_i$ by\footnote{Formally, we would need to truncate the normal distribution so that it has a support of $(0, \infty)$. However, this distinction is inconsequential in these simulations since the probability of an observation falling below 0 is extremely small (e.g., below $10^{-15}$ in all scenarios).}
\begin{equation*}
    \frac{\log2}{\lambda_i} \sim \mathrm{Normal}\left( \frac{\log 2}{0.025}, \tau^2 \right). 
\end{equation*}
We considered three values for the between-study variance $\tau^2$: 0 (no heterogeneity), 4 (moderate heterogeneity), and 12 (high heterogeneity). The moderate and high heterogeneity settings corresponded to $I^2$ values of approximately 38\% and 64\%, respectively. The censoring times were generated in the same manner as described in the group/study-level simulations. The censoring distribution of each study (i.e., uniform or exponential) was randomly assigned with equal probability.

\paragraph{Results}

\begin{figure}[H] 
  \centering
   \includegraphics[width=0.95\textwidth]{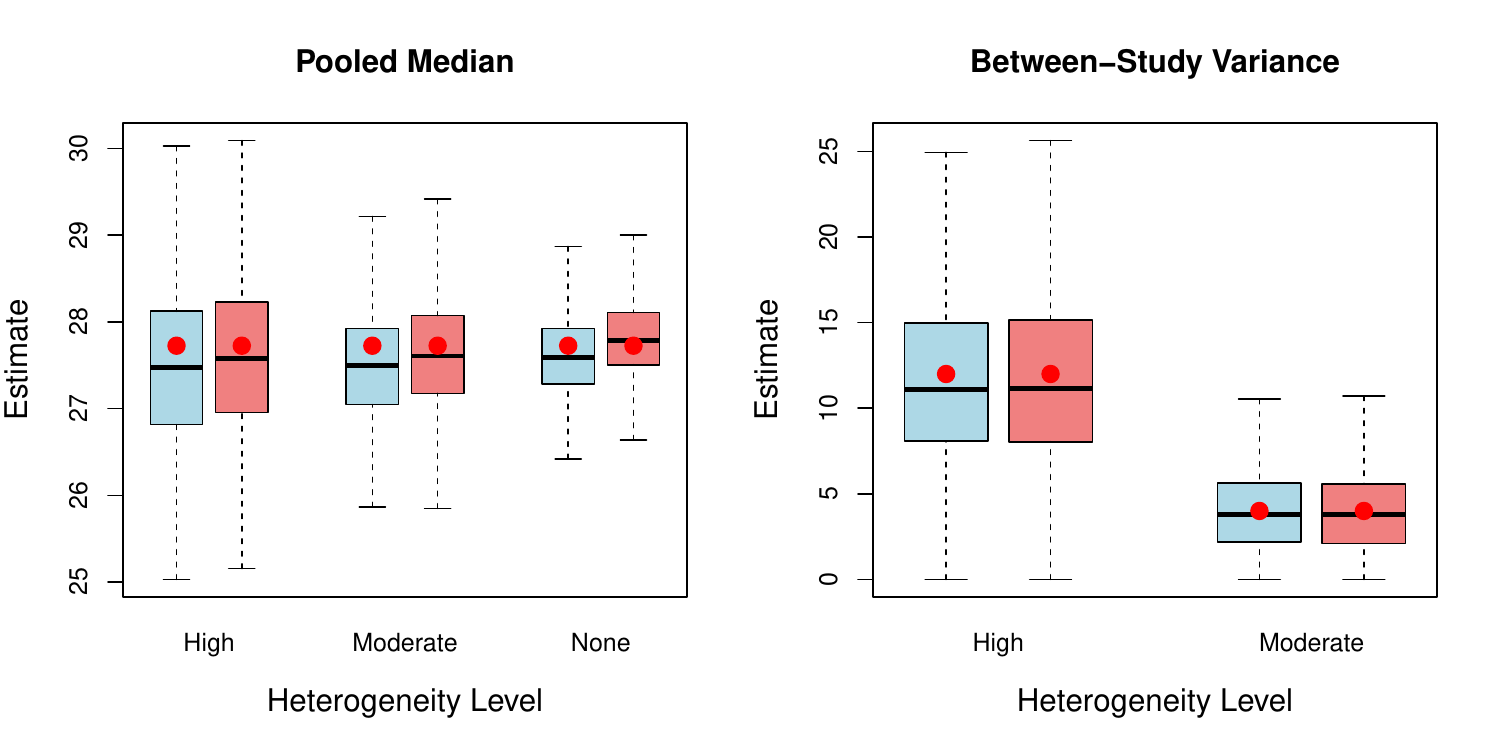}
   \caption{Estimates of the pooled median (left panel) and the between-study variance (right panel). The blue boxes correspond to the Wald approximation-based approach, and the red boxes correspond to the benchmark approach. The red dots indicate the true values} \label{fig: sim res median}
\end{figure}

The distribution of the pooled median and between-study variance estimates are illustrated in Figure \ref{fig: sim res median}. The bias, standard error, and coverage of the 95\% confidence intervals for the pooled median and between-study variance estimators are given in Table \ref{tab: sim res median}.

The Wald approximation-based approach and the benchmark approach performed very similarly to each other in each scenario. The distribution of the pooled median and between-study variance estimates were nearly identical between these two methods. In particular, both methods had low bias for estimating the pooled median and between-study variance and had similar precision for estimating these parameters. Both methods also had similar coverage probabilities for the pooled median and between-study variance, which were close to nominal levels. This suggests that suboptimal estimation of the within-study standard errors was inconsequential for the Wald approximation-based approach.

\begin{table}[H]
\caption{Simulation results for meta-analyzing median survival. The bias, standard error (SE), and coverage of the 95\% confidence intervals are reported for the Wald approximation-based approach and benchmark approach.} \label{tab: sim res median}
\begin{center}
\begin{tabular}{@{\extracolsep{6pt}}llllllll@{}}
\hline
& & \multicolumn{3}{c}{Wald Approx.\ Approach} & \multicolumn{3}{c}{Benchmark Approach} \\ \cline{3-5} \cline{6-8}
Target Parameter &  Heterogeneity & Bias & SE & Coverage & Bias & SE &  Coverage \\
  \hline
Pooled Median & High & -0.23 & 0.94 & 0.94 & -0.13 & 0.94 & 0.95 \\ 
& Moderate & -0.22 & 0.67 & 0.93 & -0.09 & 0.67 & 0.94 \\ 
& None & -0.13 & 0.47 & 0.93 & 0.08 & 0.45 & 0.96 \\  \addlinespace
Between-Study Variance & High & -0.26 & 5.42 & 0.95 & -0.11 & 5.67 & 0.95 \\ 
& Moderate & 0.15 & 2.64 & 0.96 & 0.08 & 2.62 & 0.95 \\ 
   \hline
\end{tabular}
\end{center}
\end{table}

\subsubsection{Meta-analysis of the difference of median survival} \label{sec: sim median difference}

\paragraph{Data generation}

We generated longitudinal data of two-group primary studies in a similar manner as that for one-group primary studies in Section \ref{sec: sim median}. The sample sizes of the two groups were set to be equal and were simulated from a discrete $\mathrm{Uniform}(50, 1000)$ distribution. We generated the underlying event times in group 1 in the same manner as the simulation design in Section \ref{sec: study level}, i.e., from exponential distributions with varying $\lambda_i$ to create between-study heterogeneity. We generated the underlying event times in group 2 from an $\mathrm{Exponential}(0.025)$ distribution. This implies that the true study-specific differences of median survival times follow a normal distribution with mean 0 and variance $\tau^2$ ($\tau^2 \in \{0, 4, 12\}$). The moderate and high heterogeneity settings corresponded to $I^2$ values of approximately 23\% and 47\%, respectively. The censoring distribution in both groups (i.e., uniform or exponential) was randomly assigned with equal probability.

\paragraph{Results}

\begin{figure}[H] 
  \centering
   \includegraphics[width=0.95\textwidth]{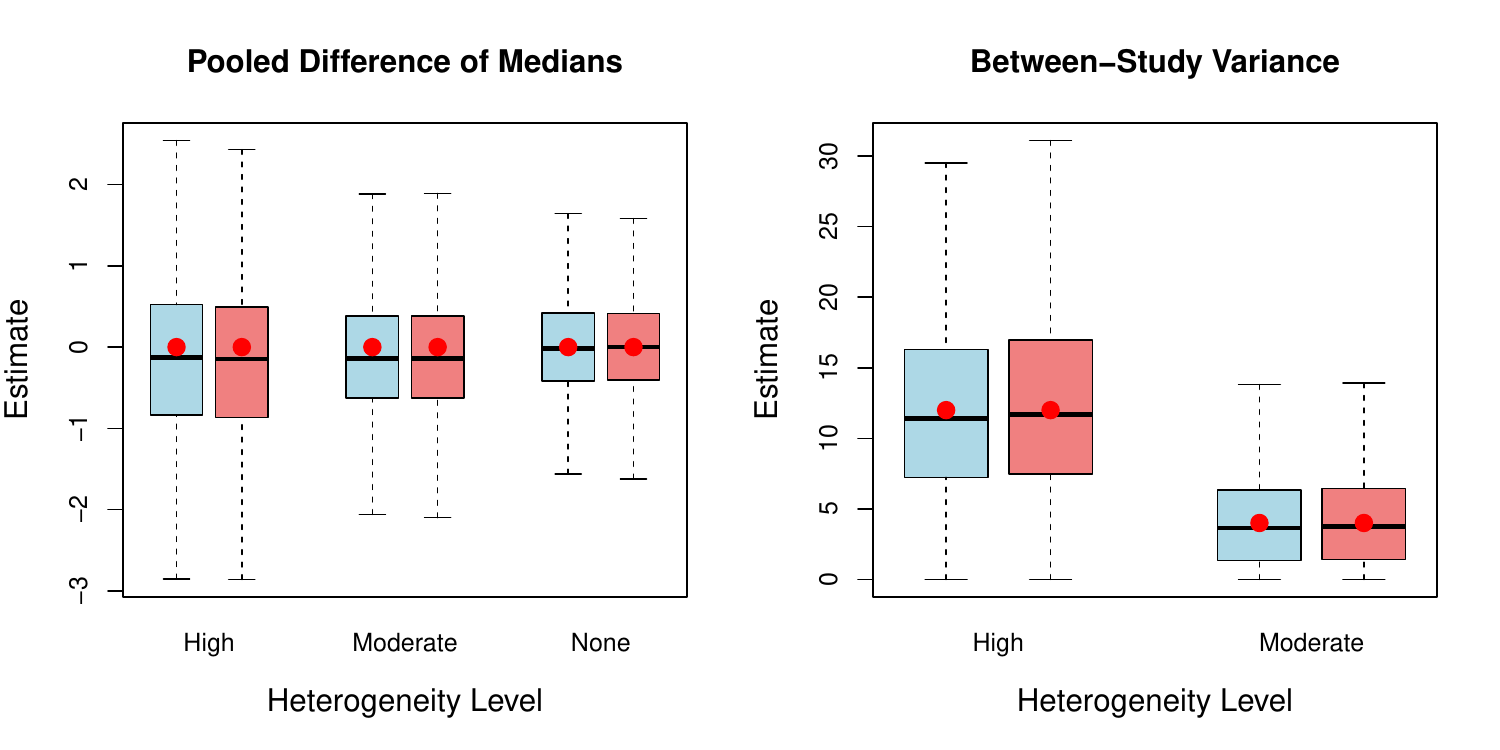}
   \caption{Estimates of the pooled difference of medians (left panel) and the between-study variance (right panel). The blue boxes correspond to the Wald approximation-based approach, and the red boxes correspond to the benchmark approach. The red dots indicate the true values} \label{fig: sim res median difference}
\end{figure}

Figure \ref{fig: sim res median difference} illustrates the distribution of the estimates of the pooled difference of medians and between-study variance in each scenario. Table \ref{tab: sim res median difference} lists the bias, standard error, and coverage of the 95\% confidence intervals for these parameters. 

Similar to the simulations for meta-analyzing median survival, we found that the Wald approximation-based approach performed very similarly to the benchmark approach. Both methods had nearly identical values for the bias, standard error, and coverage of their 95\% confidence intervals for the pooled difference of medians and the between-study variance in each scenario. Their confidence intervals for the pooled difference of medians and between-study variance had nominal coverage or slight overcoverage in all scenarios. The slight overcoverage of the confidence interval for the pooled difference of medians may be due to using Hartung-Knapp adjustment in the random effects scenarios and t-interval in the common effect scenario.

\begin{table}[H]
\caption{Simulation results for meta-analyzing the difference of median survival. The bias, standard error (SE), and coverage of the 95\% confidence intervals are reported for the Wald approximation-based approach and benchmark approach.} \label{tab: sim res median difference}
\begin{center}
\begin{tabular}{@{\extracolsep{6pt}}llllllll@{}}
\hline
& & \multicolumn{3}{c}{Wald Approx.\ Approach} & \multicolumn{3}{c}{Benchmark Approach} \\ \cline{3-5} \cline{6-8}
Target Parameter & Heterogeneity & Bias & SE & Coverage & Bias & SE &  Coverage \\
  \hline
Pooled Dif.\ of Medians & High & -0.17 & 0.99 & 0.96 & -0.16 & 0.99 & 0.96 \\ 
& Moderate & -0.12 & 0.79 & 0.95 & -0.12 & 0.78 & 0.96 \\ 
& None & -0.01 & 0.61 & 0.97 & -0.00 & 0.60 & 0.97 \\    \addlinespace
Between-Study Variance & High & 0.18 & 6.85 & 0.95 & 0.56 & 7.00 & 0.95 \\ 
& Moderate & 0.21 & 3.58 & 0.97 & 0.38 & 3.75 & 0.96 \\ 
   \hline
\end{tabular}
\end{center}
\end{table}

\subsubsection{Meta-analysis of the ratio of median survival} \label{sec: sim median ratio}

\paragraph{Data generation}

We generated longitudinal data for two-group primary studies in a very similar manner as that in the previous subsection. The only difference was in how the parameter $\lambda_i$ was generated. Here, we generated $\lambda_i$ so that the true study-specific \textit{log ratios of medians} follow a normal distribution with mean 0 and variance $\tau^2$. Specifically, we generated $\lambda_i$ by
\begin{equation*}
    \log \left( \frac{0.025}{\lambda_i} \right) \sim \mathrm{Normal}( 0, \tau^2).
\end{equation*}
We considered three values of $\tau^2$: 0 (no heterogeneity), 1/100 (moderate heterogeneity), and 3/100 (high heterogeneity). The moderate and high heterogeneity settings corresponded to $I^2$ values of approximately 38\% and 64\%, respectively.

\paragraph{Results}

\begin{figure}[H] 
  \centering
   \includegraphics[width=0.95\textwidth]{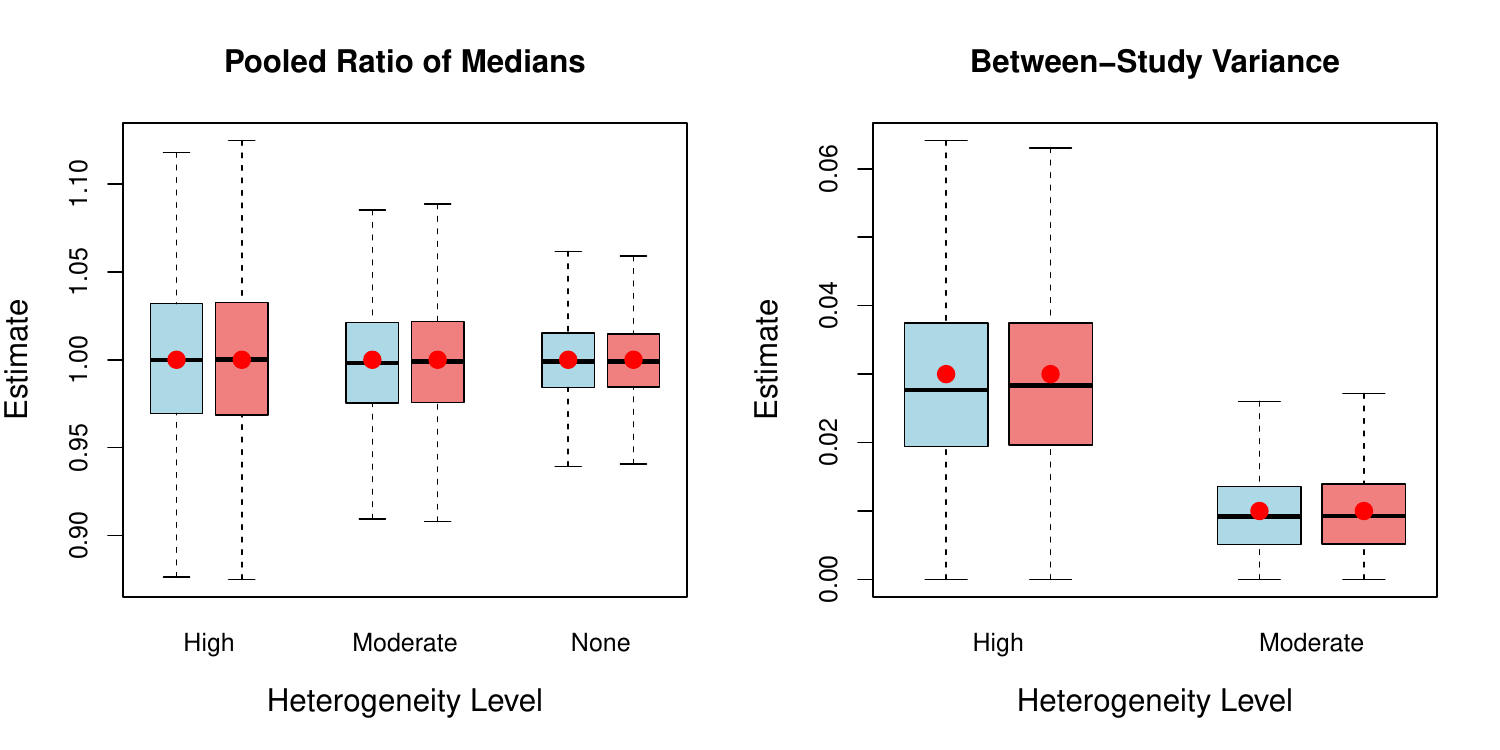}
   \caption{Estimates of the pooled ratio of medians (left panel) and the between-study variance (right panel). The blue boxes correspond to the Wald approximation-based approach, and the red boxes correspond to the benchmark approach. The red dots indicate the true values} \label{fig: sim res median ratio}
\end{figure}

Similar to the simulation studies in the previous two subsections, we summarize the simulation results in Figure \ref{fig: sim res median ratio} and Table \ref{tab: sim res median ratio}. Once again, the Wald approximation-based approach performed very similarly to the benchmark approach. In each scenario, these methods had similar bias, standard error, and confidence interval coverage for the pooled ratio of medians and between-study variance. The confidence intervals also had nominal or near-nominal coverage in each scenario.

\begin{table}[H]
\caption{Simulation results for meta-analyzing the ratio of median survival. The bias, standard error (SE), and coverage of the 95\% confidence intervals are reported for the Wald approximation-based approach and benchmark approach. The bias and standard error values are multiplied by 100.} \label{tab: sim res median ratio}
\begin{center}
\begin{tabular}{@{\extracolsep{6pt}}llllllll@{}}
\hline
& & \multicolumn{3}{c}{Wald Approx.\ Approach} & \multicolumn{3}{c}{Benchmark Approach} \\ \cline{3-5} \cline{6-8}
Target Parameter & Heterogeneity & Bias & SE & Coverage & Bias & SE &  Coverage \\
  \hline
Pooled Ratio of Medians & High & 0.06 & 4.78 & 0.94 & 0.07 & 4.78 & 0.94 \\ 
& Moderate & -0.10 & 3.33 & 0.95 & -0.07 & 3.32 & 0.95 \\ 
& None & -0.06 & 2.28 & 0.96 & -0.04 & 2.26 & 0.96 \\    \addlinespace
Between-Study Variance & High & -0.06 & 1.35 & 0.96 & -0.04 & 1.35 & 0.96 \\ 
& Moderate & 0.00 & 0.67 & 0.95 & 0.00 & 0.66 & 0.96 \\ 
   \hline
\end{tabular}
\end{center}
\end{table}

\subsubsection{Impact of misspecifying the form of between-study heterogeneity} \label{sec: sim misspecified}

The meta-analysis of the difference of medians and the meta-analysis of the ratio of medians methods involve different assumptions on the form of the between-study heterogeneity. The difference of medians meta-analysis assumes heterogeneity on the additive scale (i.e., the true study-specific differences of medians are normally distributed) and the ratio of medians meta-analysis assumes heterogeneity on the multiplicative scale (i.e., the true study-specific log ratios of medians are normally distributed). Consequently, when applying both of these methods to the same data (e.g., as in Section \ref{sec: application}), one may be concerned about misspecifying the form of between-study heterogeneity in at least one of these methods. In Section 5.1 of the Supplementary Material, we evaluate the performance of these two methods in our simulation study when misspecifying the form of the between-study heterogeneity in this manner. We found that such misspecification had little impact on the performance of these methods. Related, see \cite{jackson2018should} for a discussion of implications of the distributional assumptions on between-study heterogeneity.

\subsubsection{Impact of small within-study sample sizes}

We additionally performed analogous simulations where the within-study sample sizes were considerably smaller. As the study-level simulations found that the Wald approximation can perform poorly when sample sizes are small, these simulations illustrate the downstream consequences of applying the Wald approximation in such meta-analyses. Specifically, we re-performed the simulation studies of the meta-analysis of median survival (Section \ref{sec: sim median}), difference of median survival (Section \ref{sec: sim median difference}), and ratio of median survival  (Section \ref{sec: sim median ratio}), where the sample sizes were drawn from a discrete $\mathrm{Uniform}(20, 50)$ distribution rather than a discrete $\mathrm{Uniform}(50, 1000)$ distribution. Detailed results are given in Section 5.2 of the Supplementary Material.

For all three simulation studies, the Wald approximation-based approach often remained unbiased for estimating the pooled outcome measure and generally had similar efficiency as the benchmark approach. However, the Wald approximation-based approach had below nominal coverage of its 95\% confidence intervals for the pooled outcome measure in the common effect scenarios (e.g., dropping to 0.65 for the pooled median). The Wald approximation-based approach also generally overestimated the between-study variance and had below nominal coverage for its 95\% confidence intervals for this target parameter. Both of these issues can be seen as consequences of the Wald approximation systematically underestimating the within-study variances, as found in the study-level simulations with small sample sizes.

\section{Application} \label{sec: application}

In this section, we illustrate an application of the Wald approximation-based approach to a recent meta-analysis which investigated differences in survival between individuals randomized to experimental groups versus comparator groups in randomized trials for cancer therapies \cite{iskander2024benefits}. For the purposes of this illustration, we focus on randomized trials in non-small cell lung cancer (NSCLC) and consider the outcome of overall survival (OS). The data and code used for this application are available at  \url{https://github.com/stmcg/ma-median-survival}.

A total of 28 studies reported estimates of median OS and 95\% confidence intervals separately in the experimental and comparator groups (see Section 6 of the Supplementary Material). The average sample size in the experimental group was 228 and the average sample size in the comparator group was 205. Differences of median OS ranged from -5.39 months (favoring comparator group) to 12.40 months (favoring experimental group). Ratios of median OS ranged from 0.57 (favoring comparator group) to 1.82 (favoring experimental group).

We applied the Wald approximation-based approach to meta-analyze the median OS in the comparator group, difference of median OS between the experimental and comparator groups, and ratio of median OS between the experimental and comparator groups. Although substantive interest primarily lies in quantifying contrasts in OS between the two groups (e.g., differences or ratios of medians), we included the analysis of the median OS in the control arm to simply illustrate the approach applied to estimate a pooled median. We used random effects models and estimated the between-study variance by REML, as in the simulation studies. We assessed between-study heterogeneity based on the $I^2$ index \cite{higgins2002quantifying, higgins2003measuring} and 95\% prediction intervals (PIs) \cite{higgins2009re, inthout2016plea}.

The pooled median survival estimate in the comparator group was 12.81 months (95\% CI: 10.85, 14.77). As one may expect, there was a substantial amount of between-study heterogeneity. The 95\% prediction interval was wide (2.85 to 22.77), and the estimated $I^2$ index was 95.03\%.

\begin{figure} [H]
    \centering
    \includegraphics[width=0.95\textwidth]{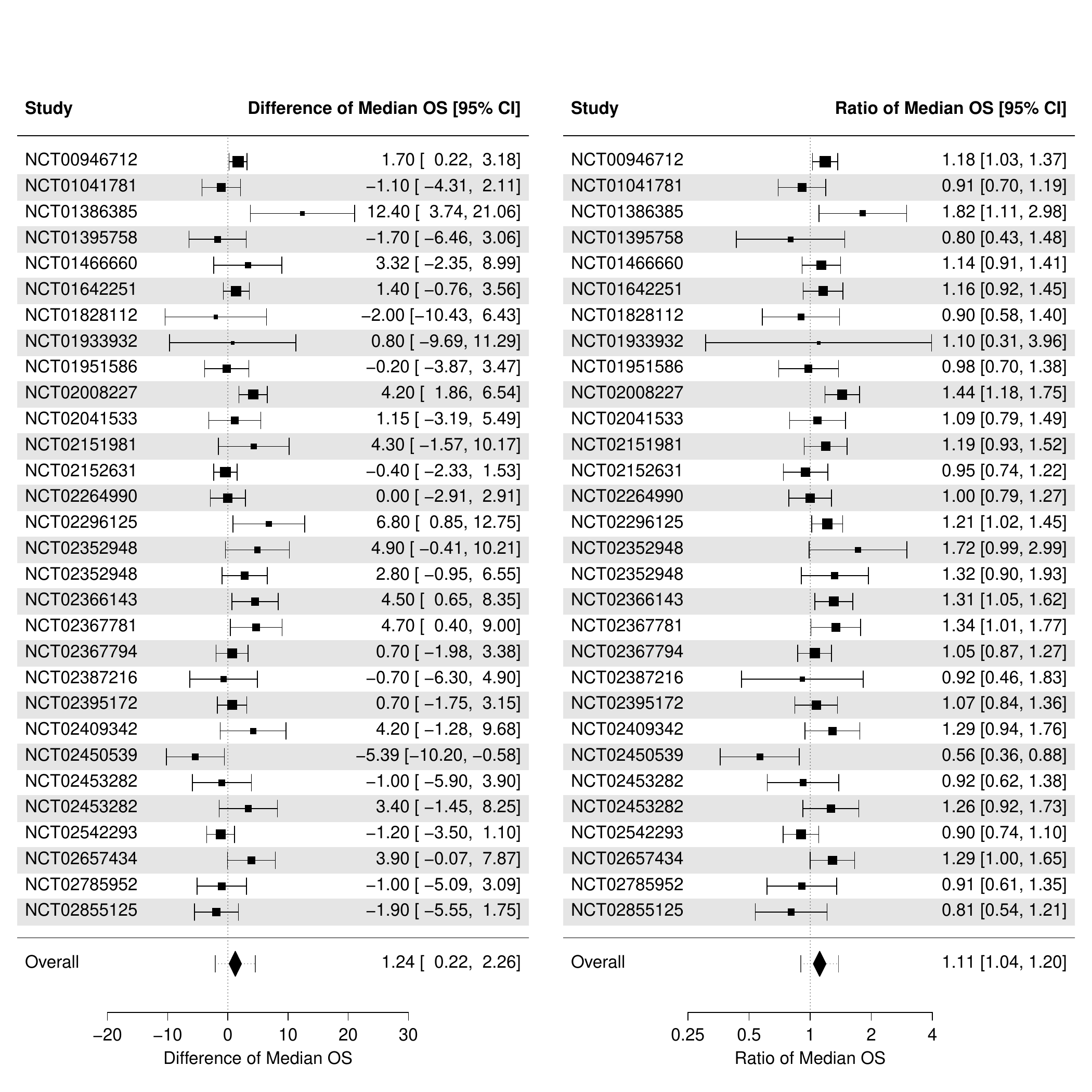}
    \caption{Forest plots for the difference of median overall survival (OS) (left panel) and ratio of median overall survival (right panel) between individuals randomized to experimental vs comparator arms in non-small cell lung cancer (NSCLC) trials. The dashed lines around the pooled estimate reflect the 95\% prediction interval\label{fig:forest both}}
\end{figure}

The forest plots for the difference of medians and ratio of medians analyses are given in Figure \ref{fig:forest both}. The pooled difference of medians estimate was 1.24 months (95\% CI: 0.22, 2.26), suggesting a small survival advantage for the experimental group.  The pooled ratio of medians estimate was 1.11 (95\% CI: 1.04, 1.20), also suggesting a relatively small advantage for the experimental group. Both of these meta-analyses had a moderate amount of between-study heterogeneity. The difference of medians analysis had a 95\% prediction interval ranging from -2.09 to 4.57 months and had an estimated $I^2$ index of 44.91\%. The ratio of medians analysis had a 95\% prediction interval ranging from 0.90 to 1.38 and had an estimated $I^2$ index of 35.56\%.

Two of the 28 studies (NCT02352948 and NCT02453282) each contributed two sets of estimates to our meta-analysis. NCT02352948 consisted of two independent sub-studies, and NCT02453282 included two independent experimental groups. Similar results were obtained when using multivariate meta-analytic approaches that account for potential correlation arising from these two studies providing multiple estimates (e.g., multivariate models with study-specific random intercepts). For simplicity and consistency with the rest of the paper, we presented the univariate analysis here.

\section{Discussion}

In this article, we present an inverse-variance weighted approach to meta-analyzing median survival times when primary studies report Kaplan-Meier estimates of median survival times and confidence intervals. In particular, we described methods to meta-analyze the median survival time, the difference of median survival times between two groups, and the ratio of median survival times between two groups. A key aspect of these methods is that they estimate the standard errors of the study-specific median survival estimates by treating the reported confidence intervals as Wald intervals. 

Our simulation results suggest that this approach may often perform well in practice for commonly used confidence interval methods (i.e., the Brookmeyer-Crowley method and nonparametric bootstrap) when the effective sample size is reasonably large. The Wald approximation-based approach generally did not systematically over- or under-estimate the standard error of median survival in such scenarios regardless of the method used to construct the confidence interval. These results held for several different event time distributions, censoring time distributions, and sample sizes. The Wald approximation-based approach also performed well in the corresponding meta-analysis simulations. This approach was approximately unbiased for estimating the pooled outcome measure and between-study variance and had corresponding near-nominal confidence interval coverage in each setting. Furthermore, this approach was comparable to a benchmark approach that uses the actual true within-study standard errors. 

In settings where the effective sample size is small, such as due to a small number of participants and a high degree of censoring, the Wald approximation can perform poorly for estimating within-study standard errors. This can in turn lead to biased estimation of between-study heterogeneity in random effects meta-analyses and poor coverage of the pooled outcome measure in common effect meta-analyses.

\subsection{Related methods} \label{sec: literature}

Prior works have described methods for meta-analyzing median survival times. Several authors have considered methods for meta-analyzing the ratio of median survival times when confidence intervals around median survival times are not available from the primary studies \cite{simes1987confronting, michiels2005meta, vesterinen2014meta, hirst2021using}. These methods often weight studies based on their sample size. Others have explored methods for arm-wise meta-analyses of median survival times where studies are weighted based on the sample size or based on the hazard ratio confidence interval \cite{zang2013synthesis, zang2015statistical}. In these cases, estimation of between-study heterogeneity is challenging and requires strong assumptions on the distribution of the survival and censoring times (e.g., assuming that the standard error of the median survival estimator equals exactly $1/\sqrt{n}$). 

Flores and Müller \cite{flores2024clustering} proposed a Bayesian nonparametric approach to meta-analyze outcome measures based on median survival in the context of cancer immunotherapy studies. This approach assumes that the confidence intervals around the medians are constructed based on central order statistics, which can be seen as an analogous assumption as that in the Wald approximation. Poli et al.\ \cite{poli2024multivariate} proposed an alternative Bayesian nonparametric approach for meta-analyzing survival times in this context that involves a different construction for the priors. A key strength of these approaches is that they accommodate flexible patterns of heterogeneity across studies. These approaches may be an excellent option for meta-analysts who are familiar with Bayesian nonparametrics and can tailor the estimation algorithm to their application at hand. 

If Kaplan-Meier curves are also available from the primary studies, several approaches can leverage this additional information \cite{latimer2011nice}. For example, Guyot et al.\ \cite{guyot2012enhanced} developed an approach to digitally scan Kaplan-Meier curves to estimate the individual participant data. This approach has been used to meta-analyze median survival times as well as full survival curves in various settings (e.g., \cite{amonkar2022systematic, munshi2020large, syn2021association, broomfield2021life}). One may expect that approaches leveraging this additional data would perform better than the Wald approximation-based approach (e.g., their performance may be closer to the benchmark approach included in the simulations). Of course, if primary studies report estimates of the standard error of median survival, conventional inverse-variance weighted meta-analytic methods can be directly applied and may perform well (e.g., similarly to the benchmark approach).

Several methods have been developed to meta-analyze sample medians. For example, many approaches have been developed to estimate sample means and their standard errors from studies reporting sample medians and other sample quantiles (e.g., minimum and maximum values, first and third quartiles) in order to meta-analyze an outcome measure based on means (see \cite{mcgrath2024metamedian} and references within). Most of these approaches are not well-suited for survival settings because they implicitly assume that there is no censoring. Other approaches involve directly meta-analyzing sample medians (e.g., \cite{mcgrath2019one, mcgrath2020meta, ozturk2020meta, lang2023robust}), some of which do not require assuming no censoring. For example, McGrath et al. \cite{mcgrath2019one, mcgrath2020meta} described and evaluated unweighted and weighted approaches to meta-analyze the (difference of) medians based on taking the median of the (difference of) sample medians. Daniele et al. \cite{daniele2022posb309} evaluated the performance of the (weighted) median of medians method \cite{mcgrath2019one} to meta-analyze arm-wise median survival in the presence of censoring.

\subsection{Limitations and future work}

The simulation studies considered settings where primary studies report Kaplan-Meier estimates of median survival and 95\% confidence intervals based on either the Brookmeyer-Crowley method \cite{brookmeyer1982confidence} (with the log or log-minus-log transformation) or the percentile-based nonparametric bootstrap method \cite{efron1994introduction}. We chose these settings because we believe they are commonly encountered in practice for survival data. However, primary studies may use different methods for estimating median survival and/or constructing confidence intervals, in which case results may of course differ. For example, primary studies may use the Brookmeyer-Crowley method \cite{brookmeyer1982confidence} with other types of transformations, such as the logit or arcsine transformations. Studies may also report confidence levels other than 95\%, such as 80\%, 90\%, or 99\%. 

The simulation studies focused on traditional parametric event time and censoring distributions. Results may differ under more realistically complex distributions, which may be interesting to explore in future work. Moreover, to keep the results digestible, our meta-analysis simulations considered relatively limited settings (i.e., a total of 18 scenarios based on varying the outcome type, heterogeneity level, and sample sizes of the primary studies). Future work could explore the impact of a number of other factors, such as the number of primary studies, event time and censoring distributions, and random effects distribution. In our analyses, we found that the effective sample sizes of the primary studies was the strongest factor affecting the performance of the Wald approximation.

Several modifications to the methods described in this work can be considered. In settings where the Wald approximation may poorly estimate the within-study standard errors, one may consider using a jackknife approach (e.g., see Ozturk and Balakrishnan \cite{ozturk2020meta}), nonparametric bootstrap, or robust variance estimation \cite{tipton2015small, hedges2010robust} to construct a confidence interval around the pooled estimate. Such approaches are less sensitive to the study-specific weights and may yield better confidence interval coverage in these settings. These approaches may be particularly useful in settings where primary studies have small effective sample sizes, as the Wald approximation can perform poorly in such settings. Another modification that may be considered is meta-analyzing the median survival time based on a log transformation rather than on the original scale, as used for the ratio of medians meta-analysis. A key implication of meta-analyzing the untransformed or log transformed median is whether heterogeneity of the study-specific median survival times is thought to be additive or multiplicative. Based on our simulations in Section 5.1 of the Supplementary Material, we suspect that this distinction would not strongly affect the performance of the method. 

Outcome measures based on median survival have limitations that should be acknowledged \cite{uno2014moving, pak2017interpretability, mccaw2022pitfall}. Median survival is not estimable in primary studies with limited follow-up or relatively few events (unless strong assumptions are made). Further, estimators of median survival may have larger standard errors than estimators of other outcome measures. Moreover, median-based outcome measures may not capture short- or long-term survival profiles very well. For these reasons, it may also be worth considering meta-analyzing other outcome measures -- such as those based on the restricted mean survival or survival probabilities at certain follow-up time points -- if suitable data are available from the primary studies. As in any meta-analysis, providing estimates of different outcome measures can provide important insights.

\subsection{Conclusions and recommendations}

The Wald approximation-based approach may be suitable for researchers who would like to meta-analyze outcome measures based on median survival. This approach only requires that primary studies report estimates of median survival times and corresponding confidence intervals, which are conventionally reported in survival settings. The Wald approximation-based approach performed well across a wide range of scenarios in our simulations, provided that the sample sizes were not small in the primary studies. As the Wald approximation-based approach adopts an inverse-variance weighted approach, data analysts can perform a number of conventional downstream analyses in meta-analysis such as those investigating publication bias. This approach can of course be applied in more complex meta-analytic settings as well, such as in meta-regression, multivariate meta-analysis, and multilevel meta-analysis settings.

This approach is straightforward to apply for data analysts familiar with applying inverse-variance weighted meta-analysis methods (e.g., via the \textbf{metafor} \cite{viechtbauer2010conducting} or \textbf{meta} \cite{balduzzi2019} R packages). We also provide software that directly implements the Wald approximation-based approach to facilitate its application. The \texttt{metamedian\_survival} function in the \textbf{metamedian} R package \cite{mcgrath2024metamedian} (starting from version 1.2.0) implements the Wald approximation-based approach to meta-analyze medians, difference of medians, and ratios of medians.

\section*{Acknowledgements}
We would like to thank Renata Iskander for helping collect the data set used in the example. This work was primarily conducted while SM was at Harvard University. The simulations in this work were run on the FASRC Cannon cluster supported by the FAS Division of Science Research Computing Group at Harvard University.

\section*{Financial disclosure}

The authors declare that no specific funding has been received for this article.

\section*{Conflict of interest}

The authors declare no potential conflict of interests.

\section*{Open Research}
The data and code used in the simulation study and data application are available at  \url{https://github.com/stmcg/ma-median-survival}.

\bibliographystyle{unsrt}
\bibliography{references}

\end{document}

\maketitle
\tableofcontents

\newpage

\section{Review of Kaplan-Meier estimation} \label{appendix: km}

In this section, we briefly review aspects of Kaplan-Meier estimation that are used throughout the main text. Proofs of all results can be found in Andersen \cite{andersen1993statistical}.

\subsection{Setup and notation}

We adopt the notation from Section 2.2.1.2 of the main text. That is, we let $T_1, \dots, T_n$ denote i.i.d.\ event times with survival function $S(t)$, density $f(t)$, and distribution function $F(t) = 1 - S(t)$. Let $C_1, \dots, C_n$ denote i.i.d.\ censoring times independent of the survival times, with observed data $Y_k = \min(T_k, C_k)$ and event indicator $\delta_k = I(T_k \leq C_k)$ for $k = 1, \ldots, n$.

Let $0 < t_1 < \cdots < t_{r}$ denote the observed event times. For each observed event time $t_i$, let $d_i$ denote the number of events at time $t_i$ and $n_i$ denote the number of individuals at risk prior to time $t_i$.

Throughout this section, we let $t \in [0, t^\ast)$ where $t^\ast = \sup \{t: S(t) > 0\}$.

\subsection{Estimation of the survival function}

The Kaplan-Meier estimator of $S(t)$ is given by
\begin{equation*}
    \widehat{S}_{n}(t) = \prod_{i: t_i \leq t} \left( 1- \frac{d_i}{n_i} \right).
\end{equation*}
Under the regularity conditions given in Section 2.2.1.2 of the main text, the Kaplan-Meier estimator of $S(t)$ is consistent and asymptotically normal with 
\begin{equation*}
    \sqrt{n}(\widehat S_n(t) - S(t)) \xrightarrow{d} \text{Normal}(0, \sigma^2(t)).
\end{equation*}
The variance of $\widehat{S}_{n}(t)$ can be estimated by the Greenwood formula:
\begin{equation*}
    \widehat{\Var}(\widehat{S}_{n}(t))  = \widehat{S}_{n}^2(t) \sum_{i: t_i \leq t} \frac{d_i}{n_i (n_i - d_i)}.
\end{equation*}
Throughout, we let $\hat{\sigma}^2_n(t)$ denote the estimator of $\sigma^2(t)$ based on the Greenwood formula scaled by $n$.

One can construct Wald-type confidence intervals for $S(t)$ based on these estimators. For example, a $100(1-\alpha)\%$ confidence interval is given by
\begin{equation*}
    \widehat{S}_{n}(t) \pm z_{1-\alpha/2} \sqrt{\widehat{\Var}(\widehat{S}_{n}(t))}
\end{equation*}
where $z_{1-\alpha/2}$ denotes the $1 - \alpha/2$ quantile of the standard normal distribution. Since $S(t) \in [0, 1]$, one often applies a transformation (e.g., log or log-minus-log) to construct the confidence interval for better small sample properties. A $100(1-\alpha)\%$ confidence interval based on the log transformation is given by
\begin{equation*}
    \widehat{S}_{n}(t) \exp \left[ \pm z_{1-\alpha/2} \frac{\sqrt{\widehat{\Var}(\widehat{S}_{n}(t))}}{\widehat{S}_{n}(t)} \right]
\end{equation*}
and a $100(1-\alpha)\%$ confidence interval based on the log-minus-log transformation is given by
\begin{equation*}
    \widehat{S}_{n}(t)^{\exp \left[ \pm z_{1-\alpha/2} \frac{\sqrt{\widehat{\Var}(\widehat{S}_{n}(t))}}{\widehat{S}_{n}(t)\log \widehat{S}_{n}(t)} \right]}.
\end{equation*}

\subsection{Estimation of quantiles}
We next consider estimation of quantiles of the distribution of event times. The $p$th quantile and the Kaplan-Meier estimator of the $p$th quantile are defined by
\begin{align*}
    \xi_p & = \inf \{t: S(t) \leq 1 - p \} \\
    \hat{\xi}_p & = \inf \{t: \widehat{S}(t) \leq 1 - p \}.
\end{align*}
For the median survival time, we have $p = 1/2$ and thus $\xi_{1/2} = m$ and $\hat{\xi}_{1/2} = \widehat{m}_n$, using the notation from the main text.

Under the regularity conditions given in Section 2.2.1.2 of the main text, $\hat{\xi}_p$ is consistent and asymptotically normal with
\begin{equation*}
    \sqrt{n}(\hat{\xi}_p - \xi_p) \xrightarrow{d} N\left(0, \frac{\sigma^2(\xi_p)}{f^2(\xi_p)}\right).
\end{equation*}
The variance of $\hat{\xi}_p$ can be estimated by
\begin{equation*}
    \widehat{\Var}(\hat{\xi}_p) = \frac{\hat{\sigma}^2(\hat{\xi}_p)}{n \hat{f}^2(\hat{\xi}_p)}
\end{equation*}
where $\hat{f}$ may be based on a kernel function estimator of the density $f$, for example.

One can construct a $100(1-\alpha)\%$ Wald-type confidence interval for $\xi_p$ by
\begin{equation*}
    \hat{\xi}_p \pm z_{1-\alpha/2} \sqrt{\widehat{\Var}(\hat{\xi}_p)}.
\end{equation*}
Since $\xi_p \in [0, t^\ast]$, one may consider applying a transformation such as the log transformation for better small sample properties. However, a disadvantage for Wald-type confidence intervals for $\hat{\xi}_p$ is that they require an estimate of the probability density function $f$. 

A more commonly used approach to construct a $100(1-\alpha)\%$ confidence interval for $\xi_p$ is the Brookmeyer-Crowley method. This approach inverts a hypothesis test of the null $\xi_p = \xi_p^0$ versus the alternative $\xi_p \neq \xi_p^0$ based on the asymptotic normality of $\widehat{S}_{n}(\xi_p^0)$ (or $g(\widehat{S}_{n}(\xi_p^0))$ for some transformation $g$). That is, the interval is the set of all $\xi_p^0$ such that one fails to reject the null hypothesis when testing $\xi_p = \xi_p^0$ versus $\xi_p \neq \xi_p^0$ at the $\alpha$ significance level. More formally, the $100(1-\alpha)\%$ confidence interval is given by
\begin{equation*}
\left\{ \xi_p^0 \in [0, t^\ast): \frac{|g(\widehat{S}_{n}(\xi_p^0)) - g(1 - p)|}
     {| g^{\prime}(\widehat{S}_{n}(\xi_p^0))|  \hat{\sigma}_{n}(\xi_p^0)/\sqrt{n}}
\leq z_{1-\alpha/2} \right\}.
\end{equation*}
An advantage of this approach is that it does not require estimating $f$.

\section{Impact of asymmetry in confidence interval tail probabilities} \label{appendix: standard error estimators}

In this section, we provide some intuition on the impact of asymmetry in confidence interval tail probabilities (i.e., $P(l_{ij} > m_{ij})$ and $P(u_{ij} > m_{ij})$) on the Wald approximation in equation (3) in the main text. 

To fix ideas, suppose now that the confidence interval is of the form
\begin{align*}
    l_{ij} & = \widehat{m}_{ij} - k_{1, \alpha} \widehat{\mathrm{SE}}(\widehat{m}_{ij}) \\
    u_{ij} & = \widehat{m}_{ij} + k_{2, \alpha} \widehat{\mathrm{SE}}(\widehat{m}_{ij})
\end{align*}
where $k_{1, \alpha}, k_{2, \alpha}$ are unknown constants that depend on $\alpha$. Assuming that the standard error estimator is consistent, it is straightforward to see that in large samples $k_{1, \alpha} = z_{1 - \alpha_1}$ and $k_{2, \alpha} = z_{1 - \alpha_2}$ where $\alpha_1 = P(l_{ij} > m_{ij})$ and $\alpha_2 = P(u_{ij} < m_{ij})$ such that $\alpha_1 + \alpha_2 = \alpha$. Therefore, if $\alpha_1$ and $\alpha_2$ are were known, we could back-compute the standard error estimate similar to equation (3) in the main text, i.e.,
\begin{equation}
    \widehat{\mathrm{SE}}(\widehat{m}_{ij}) = \frac{u_{ij} - l_{ij}}{z_{1 - \alpha_1} + z_{1 - \alpha_2}}. \label{eq: wald two limits dif}
\end{equation}
Analogous to Remark 2.1 in the main text, one can alternatively motivate the standard error estimator in equation (\ref{eq: wald two limits dif}) when $l_{ij}$ and $u_{ij}$ are quantile estimators with $\alpha_1 = P(l_{ij} > m_{ij})$ and $\alpha_2 = P(u_{ij} < m_{ij})$.

\begin{figure} [H]
    \centering
\includegraphics[width=0.6\textwidth]{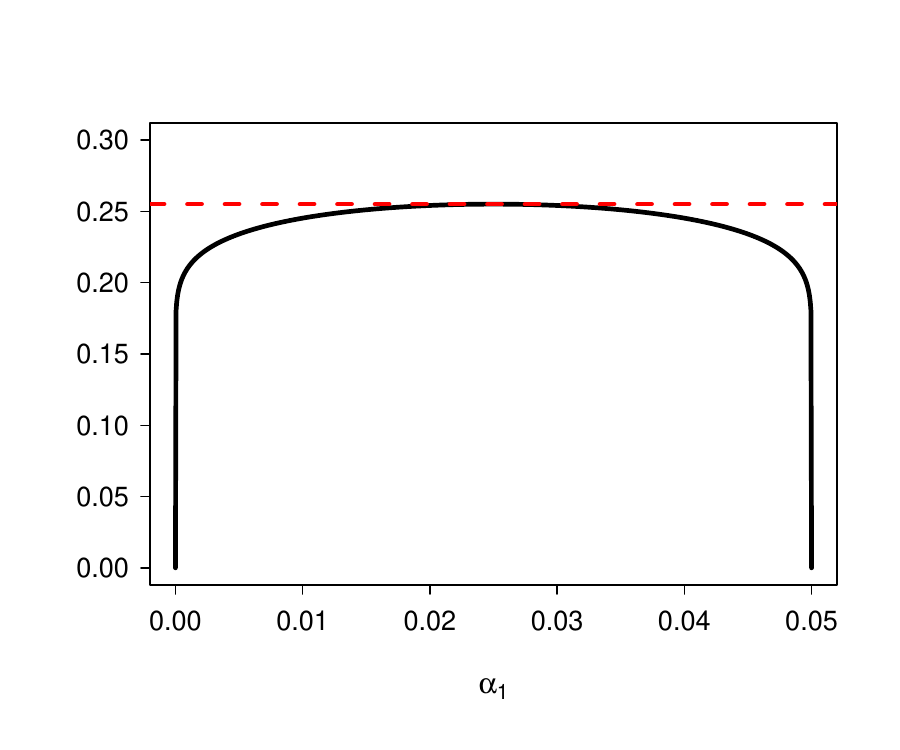}
    \caption{Comparison of $1/(z_{1 - \alpha_1} + z_{1 - \alpha_2})$ (solid black line) and $1/2z_{1-\alpha/2}$ (dashed red line) for $\alpha = 0.05$. \label{fig:intuition}}
\end{figure}

Although the values of $\alpha_1$ and $\alpha_2$ may be unknown to the meta-analyst, we can compare $1/(z_{1 - \alpha_1} + z_{1 - \alpha_2})$ (i.e., appearing in equation (\ref{eq: wald two limits dif})) and $1/2z_{1-\alpha/2}$ (i.e., appearing in equation (3) in the main text) across a range of $\alpha_1$ and $\alpha_2$ values. One can see that $1/(z_{1 - \alpha_1} + z_{1 - \alpha_2})$ reaches a maximum of $1/2z_{1-\alpha/2}$ when $\alpha_1 = \alpha_2 = \alpha/2$. Furthermore, unless $\alpha_1$ is nearly equal to 0 or $\alpha$ (and thus $\alpha_2$ is nearly $\alpha$ or 0), $1/(z_{1 - \alpha_1} + z_{1 - \alpha_2})$ is close to its maximum of $ 1 / 2 z_{1-\alpha/2}$. We illustrate this in Figure \ref{fig:intuition} and Table \ref{tab: wald clarification} for $\alpha=0.05$, noting that similar conclusions hold for other common $\alpha$ levels.

In summary, when confidence intervals are asymmetric in their tail probabilities, the Wald approximation may overestimate the standard error of median survival and the bias may increase as the degree of asymmetry in the confidence interval increases. However, the magnitude of the bias may often be small unless one of the tail probabilities is nearly 0.

\begin{table}[H]
\caption{Examples of the impact of asymmetry in confidence interval tail probabilities on the performance of the Wald approximation when $\alpha = 0.05$.} \label{tab: wald clarification}
\begin{center}
\begin{tabular}{lllll:l}
\hline
$\alpha_1$ & $\alpha_2$ & $z_{1 - \alpha_1}$ & $z_{1 - \alpha_2}$ & $1/(z_{1 - \alpha_1} + z_{1 - \alpha_2})$ & $1/(2z_{1-\alpha/2})$ \\ \hline  
0.025 & 0.025 & 1.96 & 1.96 & 0.26 & 0.26 \\ 
  0.030 & 0.020 & 1.88 & 2.05 & 0.25 & 0.26 \\ 
  0.035 & 0.015 & 1.81 & 2.17 & 0.25 & 0.26 \\ 
  0.040 & 0.010 & 1.75 & 2.33 & 0.25 & 0.26 \\ 
  0.045 & 0.005 & 1.70 & 2.58 & 0.23 & 0.26 \\ 
  0.049 & 0.001 & 1.65 & 3.09 & 0.21 & 0.26 \\ 
   \hline
\end{tabular}
\end{center}
\end{table}

\section{Proof of Proposition 1}

\subsection{Preliminaries}
In addition to the notation introduced in Section 2.2.1.2 in the main text, we introduce the following notation. Define the empirical process $Z_n(t) = \sqrt{n}(\widehat S_n(t) - S(t))$. Standard martingale theory establishes that $Z_n(t)$ converges weakly to a zero-mean Gaussian process $\mathcal{Z}(t)$ on a finite interval $[0, \nu]$ where $\nu > m$ (e.g., Andersen \cite{andersen1993statistical}). A key consequence of this weak convergence is the stochastic equicontinuity of $Z_n$ at $m$. That is, for any sequence of random variables $T_n \xrightarrow{p} m$:
\begin{equation} \label{eq:equicontinuity}
|Z_n(T_n) - Z_n(m)| \xrightarrow{p} 0.
\end{equation}

We use standard stochastic order notation where  $X_n = o_p(a_n)$ means $X_n/a_n \xrightarrow{p} 0$, and $X_n = O_p(a_n)$ means $X_n/a_n$ is bounded in probability.

Recall that we consider the $1-\alpha$ test-inversion confidence interval $[L_n, U_n]$ for the median, defined as
\begin{equation*}
\left\{ t : \frac{|\widehat S_n(t) - 1/2|}{\hat \sigma_n(t)/\sqrt{n}} \le z_{1-\alpha/2} \right\}
\end{equation*}
where $\hat{\sigma}_n^2(t)$ is based on the scaled Greenwood's formula: $\hat \sigma_n^2(t) = n \cdot \widehat{\text{Var}}(\widehat S_n(t))$. Under the regularity conditions given in the main text, $\hat \sigma_n^2(t)$ is uniformly consistent for $\sigma^2(t)$ in a neighborhood of $m$ \cite{andersen1993statistical}.

\subsection{Proof}
By the definition of the Brookmeyer-Crowley interval, $L_n$ is the largest value satisfying:
\begin{equation} \label{eq:L_implicit}
\frac{\widehat{S}_n(L_n) - 1/2}{\hat{\sigma}_n(L_n)/\sqrt{n}} = z_{1-\alpha/2}
\end{equation}
up to the discretization (jump size) imposed by the step function $\widehat{S}_n$. Since the jump size is $O_p(n^{-1})$, which is negligible compared to the $O_p(n^{-1/2})$ precision required here, we have:
\begin{equation} \label{eq:boundary_L}
\widehat S_n(L_n) = \frac{1}{2} + \frac{z_{1-\alpha/2}\hat \sigma_n(L_n)}{\sqrt{n}} + o_p(n^{-1/2}).
\end{equation}
Similarly, the upper limit $U_n$ satisfies:
\begin{equation} \label{eq:boundary_U}
\widehat S_n(U_n) = \frac{1}{2} - \frac{z_{1-\alpha/2}\hat \sigma_n(U_n)}{\sqrt{n}} + o_p(n^{-1/2}).
\end{equation}

First, we establish the convergence rate of $L_n$. By the functional central limit theorem for the Kaplan--Meier estimator, $Z_n = \sqrt{n}(\widehat S_n - S)$ converges weakly in $\ell^\infty[0,\nu]$ to a Gaussian process $\mathcal{Z}$. Combining this with \eqref{eq:boundary_L} and the consistency of $\hat \sigma_n$, we have:
\[
|S(L_n) - 1/2| \le |S(L_n) - \widehat S_n(L_n)| + |\widehat S_n(L_n) - 1/2| = O_p(n^{-1/2}).
\]
Since $f$ is continuous and positive at $m$, $S(t)$ is strictly decreasing with derivative bounded away from zero in a neighborhood of $m$. By the Mean Value Theorem, on the event $|L_n - m| \le \varepsilon$, we have $S(L_n) - S(m) = -f(\xi_n)(L_n - m)$ for some $\xi_n$ between $L_n$ and $m$. This implies:
\[
|L_n - m| \le \frac{2}{f(m)} |S(L_n) - 1/2| = O_p(n^{-1/2}).
\]
Thus $|L_n - m| = O_p(n^{-1/2})$. This rate allows us to specify the remainder term in the Taylor expansion of $S(L_n)$ around $m$:
\[
S(L_n) = \frac{1}{2} - f(m)(L_n - m) + o_p(n^{-1/2}).
\]

By the definition of the empirical process $Z_n$, we have the decomposition:
\[
\widehat S_n(L_n) = S(L_n) + \frac{1}{\sqrt{n}} Z_n(L_n).
\]
Applying the stochastic equicontinuity property (\ref{eq:equicontinuity}) with $T_n = L_n$ (which satisfies $L_n \xrightarrow{p} m$), we approximate the process at $L_n$ by the process at $m$:
\[
Z_n(L_n) = Z_n(m) + o_p(1) = \sqrt{n}(\widehat S_n(m) - 1/2) + o_p(1).
\]
Substituting the Taylor expansion and the process approximation back into the decomposition yields:
\begin{equation} \label{eq:L_expansion}
\widehat S_n(L_n) = \frac{1}{2} - f(m)(L_n - m) + (\widehat S_n(m) - 1/2) + o_p(n^{-1/2}).
\end{equation}

Next, we consider the variance term in (\ref{eq:boundary_L}). Since $\hat \sigma_n$ is uniformly consistent and $L_n \xrightarrow{p} m$, we have $\hat \sigma_n(L_n) \xrightarrow{p} \sigma(m)$. Thus:
\[
\frac{z_{1-\alpha/2}\hat \sigma_n(L_n)}{\sqrt{n}} = \frac{z_{1-\alpha/2}\sigma(m)}{\sqrt{n}} + o_p(n^{-1/2}).
\]
Equating (\ref{eq:L_expansion}) with the boundary condition (\ref{eq:boundary_L}):
\[
\frac{1}{2} - f(m)(L_n - m) + (\widehat S_n(m) - 1/2) = \frac{1}{2} + \frac{z_{1-\alpha/2}\sigma(m)}{\sqrt{n}} + o_p(n^{-1/2}).
\]
Solving for $L_n$:
\[
L_n = m + \frac{1}{f(m)} \left( \widehat S_n(m) - 1/2 - \frac{z_{1-\alpha/2}\sigma(m)}{\sqrt{n}} \right) + o_p(n^{-1/2}).
\]

By symmetry, for the upper limit $U_n$, equating the expansion with (\ref{eq:boundary_U}) leads to:
\[
\frac{1}{2} - f(m)(U_n - m) + (\widehat S_n(m) - 1/2) = \frac{1}{2} - \frac{z_{1-\alpha/2}\sigma(m)}{\sqrt{n}} + o_p(n^{-1/2}).
\]
Solving for $U_n$:
\[
U_n = m + \frac{1}{f(m)} \left( \widehat S_n(m) - 1/2 + \frac{z_{1-\alpha/2}\sigma(m)}{\sqrt{n}} \right) + o_p(n^{-1/2}).
\]

Finally, subtracting $L_n$ from $U_n$, the random empirical process term $(\widehat S_n(m) - 1/2)$ cancels out:
\begin{align*}
U_n - L_n &= \frac{1}{f(m)} \left[ \frac{z_{1-\alpha/2}\sigma(m)}{\sqrt{n}} - \left( - \frac{z_{1-\alpha/2}\sigma(m)}{\sqrt{n}} \right) \right] + o_p(n^{-1/2}) \\
&= \frac{2 z_{1-\alpha/2}\sigma(m)}{f(m)\sqrt{n}} + o_p(n^{-1/2}).
\end{align*}
Multiplying by $\sqrt{n} / (2 z_{1-\alpha/2})$ completes the proof.

\section{Additional materials for the study-level simulations} \label{appendix: simulations study level}

\subsection{Additional results of the primary study-level simulations}
\begin{figure} [H]
    \centering
\includegraphics[width=0.6\textwidth]{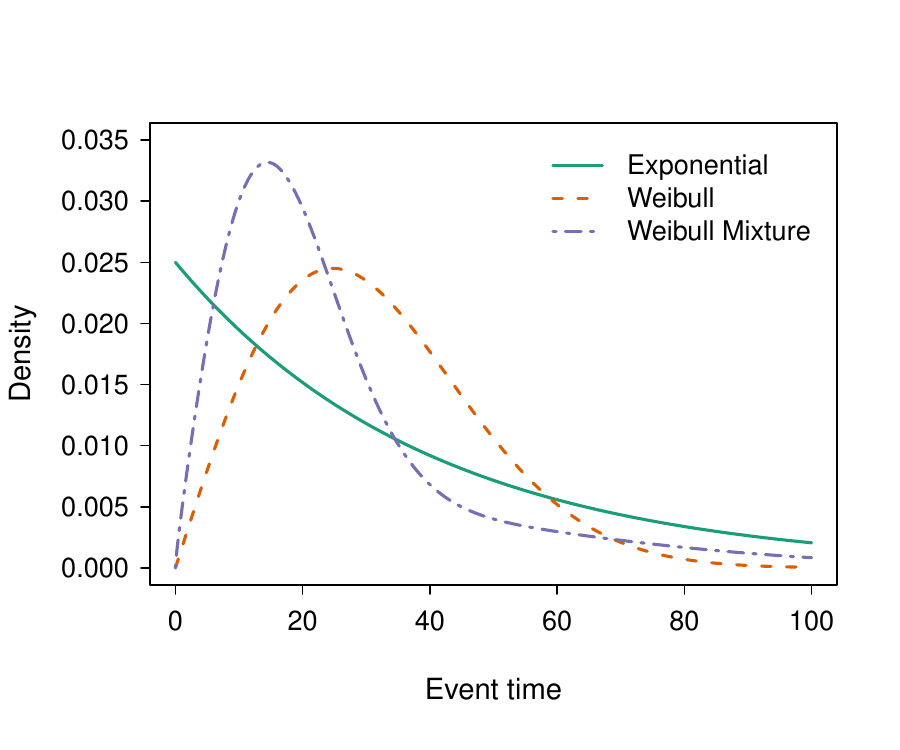}
    \caption{Probability density functions of the underlying event time distributions included in the study-level simulations.  \label{fig:distributions}}
\end{figure}

\begin{table}[H]
\caption{Percentage of censoring in the study-level simulations. The entries in the table list the percentage of individuals that were censored for each event time distribution and censoring time distribution.} \label{tab: percent censoring}
\begin{center}
\begin{tabular}{lll}
\hline
& \multicolumn{2}{c}{Censoring Time Distribution} \\ \cline{2-3}
Event Time Distribution & Uniform & Exponential  \\  \hline
Exponential & 37 & 41 \\
Weibull & 31 & 38 \\
Weibull Mixture & 26 & 33 \\ \hline
\end{tabular}
\end{center}
\end{table}

\begin{figure} [H]
    \centering
\includegraphics[width=\textwidth]{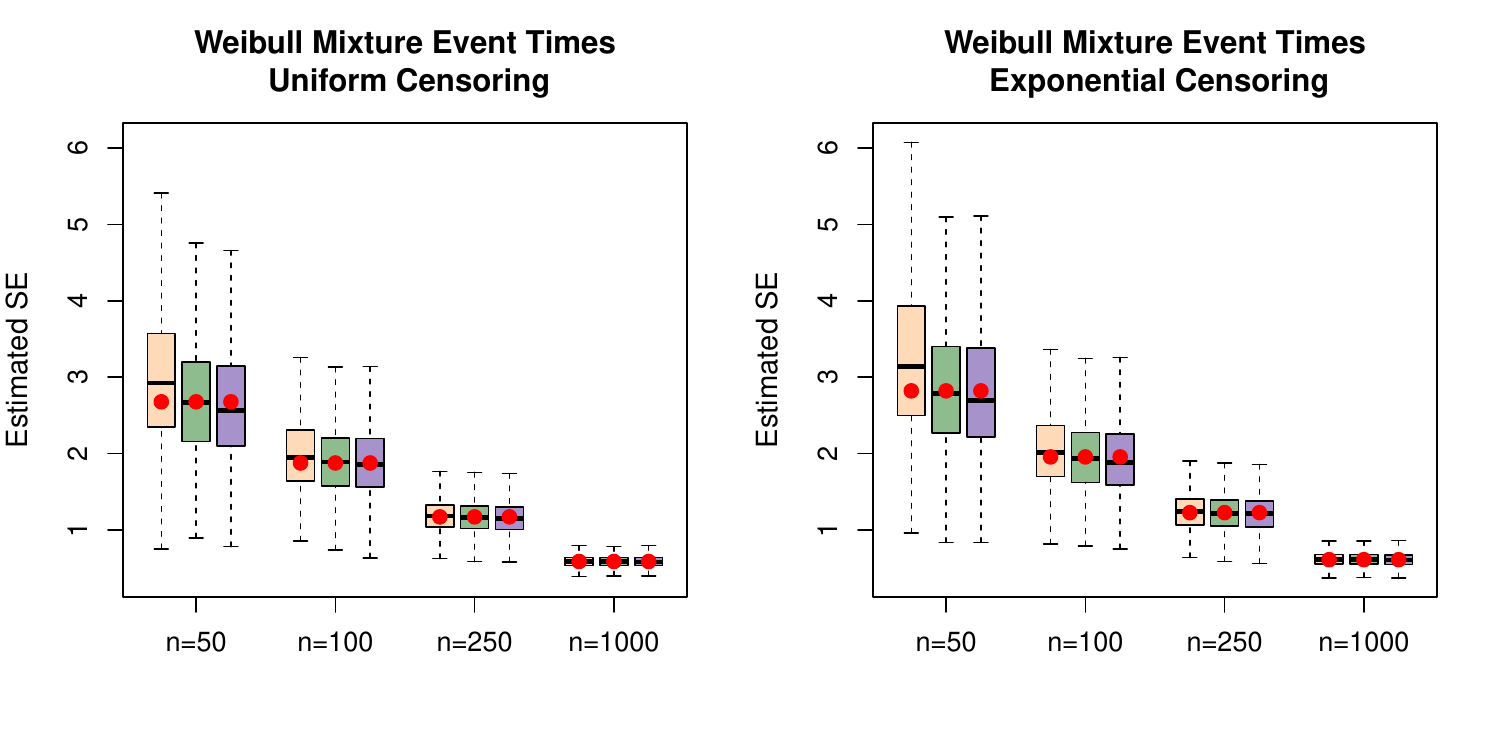}
    \caption{Study-level simulation results for the scenarios with the mixture distribution for the event times. The box plots illustrate the estimated standard errors (SEs) of the median survival time from the Wald approximation-based approach. The peach boxes correspond to when the Brookmeyer-Crowley method based on a log transformation was used to construct the 95\% confidence interval; The green boxes correspond to the Brookmeyer-Crowley method based on a log-minus-log transformation; The purple boxes correspond to the nonparametric bootstrap method. The true standard errors are illustrated by red dots.  \label{fig:sim res study-level mixture}}
\end{figure}

\begin{table}[H]
\caption{Relative bias for estimating the standard error of median survival. The ``CI Method" column describes the method used to construct the 95\% confidence interval around median survival. In this column, ``BC (log)" denotes the Brookmeyer-Crowley method with the log transformation and ``BC (log-log)" denotes the Brookmeyer-Crowley method with the log-minus-log transformation.} \label{tab: sim res study-level}
\begin{center}
\begin{tabular}{lllllll}
\hline
& & & \multicolumn{4}{c}{Sample Size} \\ \cline{4-7}
Event Time Dist. & Censoring Dist. & CI Method & $n = 50$ & $n = 100$ & $n = 250$ & $n = 1000$ \\ 
  \hline
Exponential & Uniform & BC (log) & 14 & 8 & 4 & 0 \\
 & & BC (log-log) & 4 & 2 & 1 & -1 \\ 
 & & Bootstrap & 4 & 1 & 1 & -1 \\ 
 & Exponential & BC (log) & 12 & 10 & 3 & 1 \\ 
 & & BC (log-log) & 4 & 1 & 1 & 0 \\ 
 & & Bootstrap & 3 & 1 & 0 & 0 \\ 
Weibull & Uniform & BC (log) & 11 & 3 & 0 & 0 \\ 
 &  & BC (log-log) & 4 & 0 & -1 & 0 \\ 
 &  & Bootstrap & 1 & -2 & -2 & -1 \\ 
 & Exponential & BC (log) & 12 & 5 & 2 & 1 \\ 
  &  & BC (log-log) & 2 & 1 & 1 & 1 \\ 
   &  & Bootstrap & 0 & -1 & 0 & 1 \\ 
Weibull Mixture & Uniform & BC (log) & 16 & 6 & 2 & 0 \\ 
 & & BC (log-log) & 4 & 2 & 0 & 0 \\ 
 & & Bootstrap & 2 & 1 & -1 & 0 \\ 
 & Exponential & BC (log) & 19 & 6 & 2 & 2 \\ 
  & & BC (log-log) & 4 & 1 & 0 & 1 \\ 
   & & Bootstrap & 2 & 0 & 0 & 1 \\ 
   \hline
\end{tabular}
\end{center}
\end{table}

\subsection{Impact of small sample sizes}

\begin{figure} [H]
    \centering
\includegraphics[width=\textwidth]{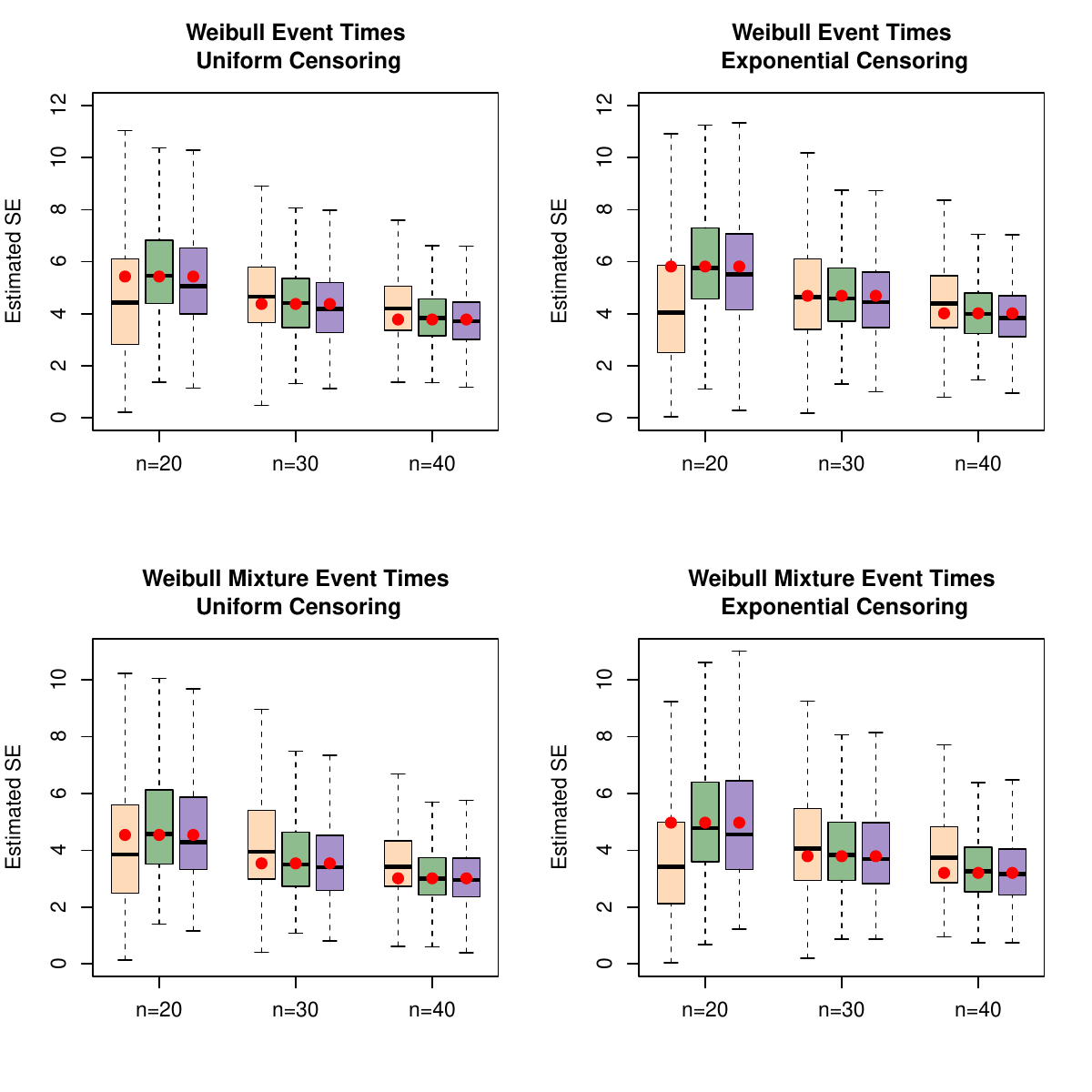}
    \caption{Study-level simulation results for the scenarios with small sample sizes and the Weibull and mixture distributions for the event times. The box plots illustrate the estimated standard errors SEs of the median survival time from the Wald approximation-based approach. The peach boxes correspond to when the Brookmeyer-Crowley method based on a log transformation was used to construct the 95\% confidence interval; The green boxes correspond to the Brookmeyer-Crowley method based on a log-minus-log transformation; The purple boxes correspond to the nonparametric bootstrap method. The true standard errors are illustrated by red dots.  \label{fig:sim res study-level mixture smalln}}
\end{figure}

\begin{table}[H]
\caption{Relative bias for estimating the standard error of median survival in the scenarios with small sample sizes. The ``CI Method" column describes the method used to construct the 95\% confidence interval around median survival. In this column, ``BC (log)" denotes the Brookmeyer-Crowley method with the log transformation and ``BC (log-log)" denotes the Brookmeyer-Crowley method with the log-minus-log transformation.} \label{tab: sim res study-level smalln}
\begin{center}
\begin{tabular}{llllll}
\hline
& & & \multicolumn{3}{c}{Sample Size} \\ \cline{4-6}
Event Time Dist. & Censoring Dist. & CI Method & $n = 20$ & $n = 30$ & $n = 40$ \\ 
  \hline
Exponential & Uniform & BC (log) & -36 & -15 & 3 \\ 
 & & BC (log-log) & -6 & -1 & 2 \\ 
 & & Bootstrap & -12 & -1 & 0 \\ 
 & Exponential & BC (log) & -39 & -22 & 1 \\ 
 & & BC (log-log) & -12 & -3 & 1 \\ 
 & & Bootstrap & -16 & -2 & 1 \\ 
Weibull & Uniform & BC (log) & -15 & 10 & 14 \\ 
 &  & BC (log-log) & 6 & 4 & 4 \\ 
 &  & Bootstrap & 0 & 0 & 0 \\ 
 & Exponential & BC (log) & -24 & 3 & 13 \\ 
  &  & BC (log-log) & 5 & 3 & 2 \\ 
   &  & Bootstrap & -1 & -1 & -1 \\ 
Weibull Mixture & Uniform & BC (log) & -6 & 24 & 22 \\ 
 & & BC (log-log) & 13 & 10 & 6 \\ 
 & & Bootstrap & 10 & 7 & 5 \\ 
 & Exponential & BC (log) & -21 & 18 & 28 \\
  & & BC (log-log) & 8 & 12 & 8 \\ 
   & & Bootstrap & 9 & 10 & 5 \\ 
   \hline
\end{tabular}
\end{center}
\end{table}

\subsection{Impact of a higher degree of censoring}

\begin{table}[H]
\caption{Percentage of censoring in the study-level simulations with a higher degree of censoring. The entries in the table list the percentage of individuals that were censored for each event time distribution and censoring time distribution.} \label{tab: percent censoring highcensoring}
\begin{center}
\begin{tabular}{lll}
\hline
& \multicolumn{2}{c}{Censoring Time Distribution} \\ \cline{2-3}
Event Time Distribution & $\mathrm{Exponential}(1/40)$ & $\mathrm{Exponential}(1/25)$  \\  \hline
Exponential & 50 & 62 \\
Weibull & 50 & 65 \\
Weibull Mixture & 43 & 56 \\ \hline
\end{tabular}
\end{center}
\end{table}

\begin{figure} [H]
    \centering
\includegraphics[width=\textwidth]{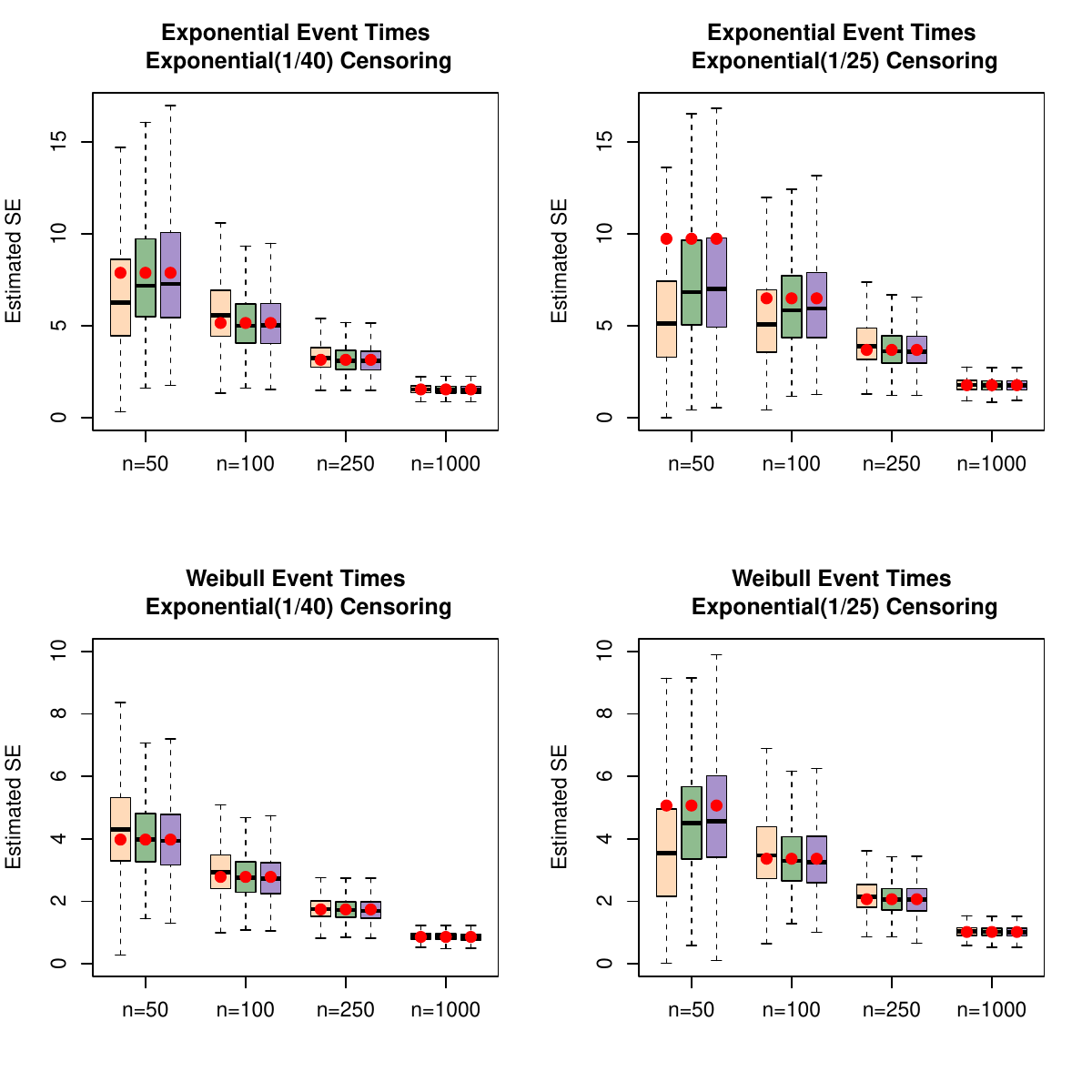}
    \caption{Study-level simulation results for the scenarios with a higher degree of censoring and the exponential and Weibull distributions for the event times. The box plots illustrate the estimated standard errors SEs of the median survival time from the Wald approximation-based approach. The peach boxes correspond to when the Brookmeyer-Crowley method based on a log transformation was used to construct the 95\% confidence interval; The green boxes correspond to the Brookmeyer-Crowley method based on a log-minus-log transformation; The purple boxes correspond to the nonparametric bootstrap method. The true standard errors are illustrated by red dots.  \label{fig:sim res study-level highcensoringpt1}}
\end{figure}

\begin{figure} [H]
    \centering
\includegraphics[width=\textwidth]{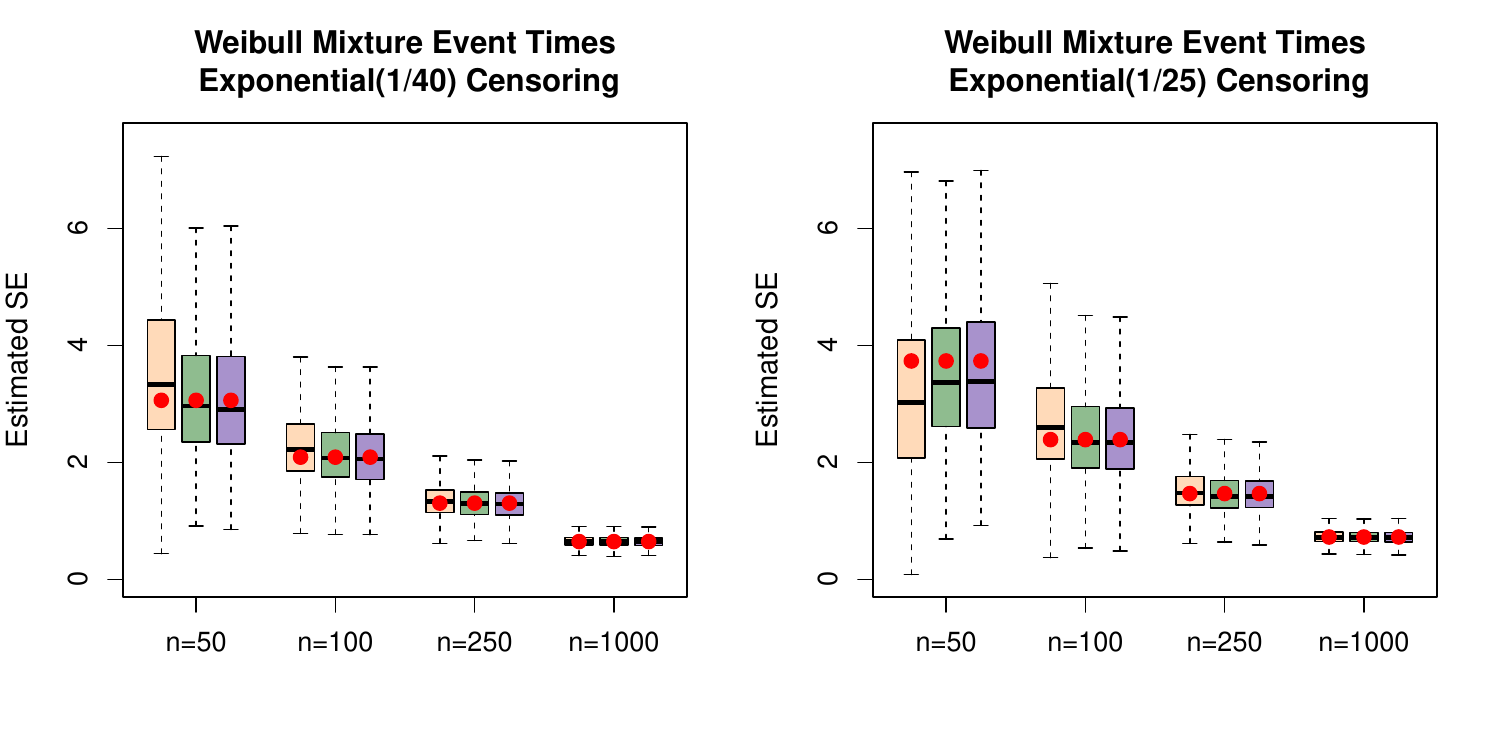}
    \caption{Study-level simulation results for the scenarios with a higher degree of censoring and the mixture distribution for the event times. The box plots illustrate the estimated standard errors SEs of the median survival time from the Wald approximation-based approach. The peach boxes correspond to when the Brookmeyer-Crowley method based on a log transformation was used to construct the 95\% confidence interval; The green boxes correspond to the Brookmeyer-Crowley method based on a log-minus-log transformation; The purple boxes correspond to the nonparametric bootstrap method. The true standard errors are illustrated by red dots.  \label{fig:sim res study-level highcensoringpt2}}
\end{figure}

\begin{table}[H]
\caption{Relative bias for estimating the standard error of median survival in the scenarios with a high degree of censoring. The ``CI Method" column describes the method used to construct the 95\% confidence interval around median survival. In this column, ``BC (log)" denotes the Brookmeyer-Crowley method with the log transformation and ``BC (log-log)" denotes the Brookmeyer-Crowley method with the log-minus-log transformation.} \label{tab: sim res study-level smalln}
\begin{center}
\begin{tabular}{lllllll}
\hline
& & & \multicolumn{4}{c}{Sample Size} \\ \cline{4-7}
Event Time Dist. & Censoring Level & CI Method & $n = 50$ & $n = 100$ & $n = 250$ & $n = 1000$ \\ 
  \hline
Exponential & Moderate & BC (log) & -14 & 14 & 6 & 1 \\
 & & BC (log-log) & 1 & 2 & 1 & 0 \\ 
 & & Bootstrap & 3 & 2 & 1 & 0 \\ 
 & High & BC (log) & -39 & -16 & 12 & 2 \\ 
 & & BC (log-log) & -21 & -4 & 4 & 0 \\ 
 & & Bootstrap & -21 & 1 & 4 & 0 \\ 
Weibull & Moderate & BC (log) & 10 & 8 & 2 & 1 \\ 
 & & BC (log-log) & 4 & 1 & 0 & 1 \\ 
 & & Bootstrap & 3 & 0 & -1 & 0 \\ 
 & High & BC (log) & -26 & 8 & 6 & 2 \\ 
 & & BC (log-log) & -6 & 2 & 1 & 1 \\ 
 & & Bootstrap & -2 & 2 & 1 & 0 \\ 
Mixture & Moderate & BC (log) & 22 & 10 & 4 & 1 \\ 
 & & BC (log-log) & 6 & 3 & 1 & 1 \\ 
 & & Bootstrap & 7 & 2 & 0 & 0 \\ 
 & High & BC (log) & -13 & 16 & 5 & 1 \\ 
 & & BC (log-log) & -1 & 5 & 1 & 1 \\ 
 & & Bootstrap & 3 & 4 & 0 & 0 \\ 
   \hline
\end{tabular}
\end{center}
\end{table}

\subsection{Impact of extreme skewness} \label{appendix: extreme skewness}

In this section, we illustrate the performance of the Wald approximation under settings with increasingly high skewness in the event time distribution.

\subsubsection{Data generation}

\begin{figure} [H]
    \centering
\includegraphics[width=0.6\textwidth]{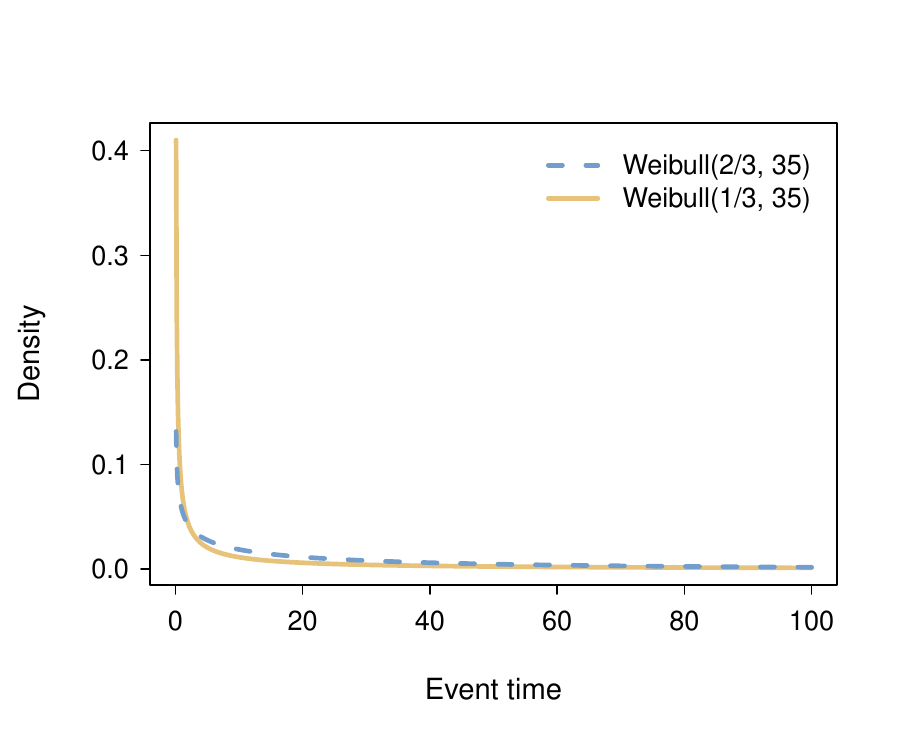}
    \caption{Probability density function of the highly skewed Weibull distributions.  \label{fig:distribution skew}}
\end{figure}

We adopt the same simulation design as in the study-level simulations in the main text (i.e., Section 3.1.1) with the following changes. Here, we generated the underlying event times from either a Weibull distribution with $(k = 2/3, \lambda=35)$ or $(k = 1/3, \lambda=35)$. Figure \ref{fig:distribution skew} illustrates the probability density functions of the two distributions. We used the distributions for censoring and used the same four sample sizes as in the main text. 

Note that decreasing the shape parameter $k$ in this context increases the skewness of the event time distribution, which can be seen in the following ways. Pearson's moment coefficient of skewness is approximately 3.8 for the $\mathrm{Weibull}(k = 2/3, \lambda=35)$ distribution and is 19.6 for the $\mathrm{Weibull}(k = 1/3, \lambda=35)$ distribution. An alternative measure of skewness called Bowley's coefficient of skewness \cite{bowley1901} has been commonly used in the context of meta-analyzing median-based outcome measures (e.g., see \cite{mcgrath2019one, mcgrath2020meta, ozturk2020meta, mcgrath2023standard, mcgrath2024metamedian}). Bowley's skewness ranges from $-1$ to $1$, where negative values indicate left skewness and positive values indicate right skewness. Bowley's skewness of the $\mathrm{Weibull}(k = 2/3, \lambda=35)$ and $\mathrm{Weibull}(k = 1/3, \lambda=35)$ distributions are approximately 0.43 and 0.77, respectively.

When the underlying event times followed the Weibull distribution with $(k = 2/3, \lambda=35)$, the percentage of censoring was 34\% in the scenarios with the uniform censoring time distribution and 38\% in the scenarios with the exponential censoring time distribution. When the underlying event times followed the Weibull distribution with $(k = 1/3, \lambda=35)$, the percentage of censoring was 36\% in the scenarios with the uniform censoring time distribution and 38\% in the scenarios with the exponential censoring time distribution.

\subsubsection{Results}

\begin{figure} [H]
    \centering
\includegraphics[width=\textwidth]{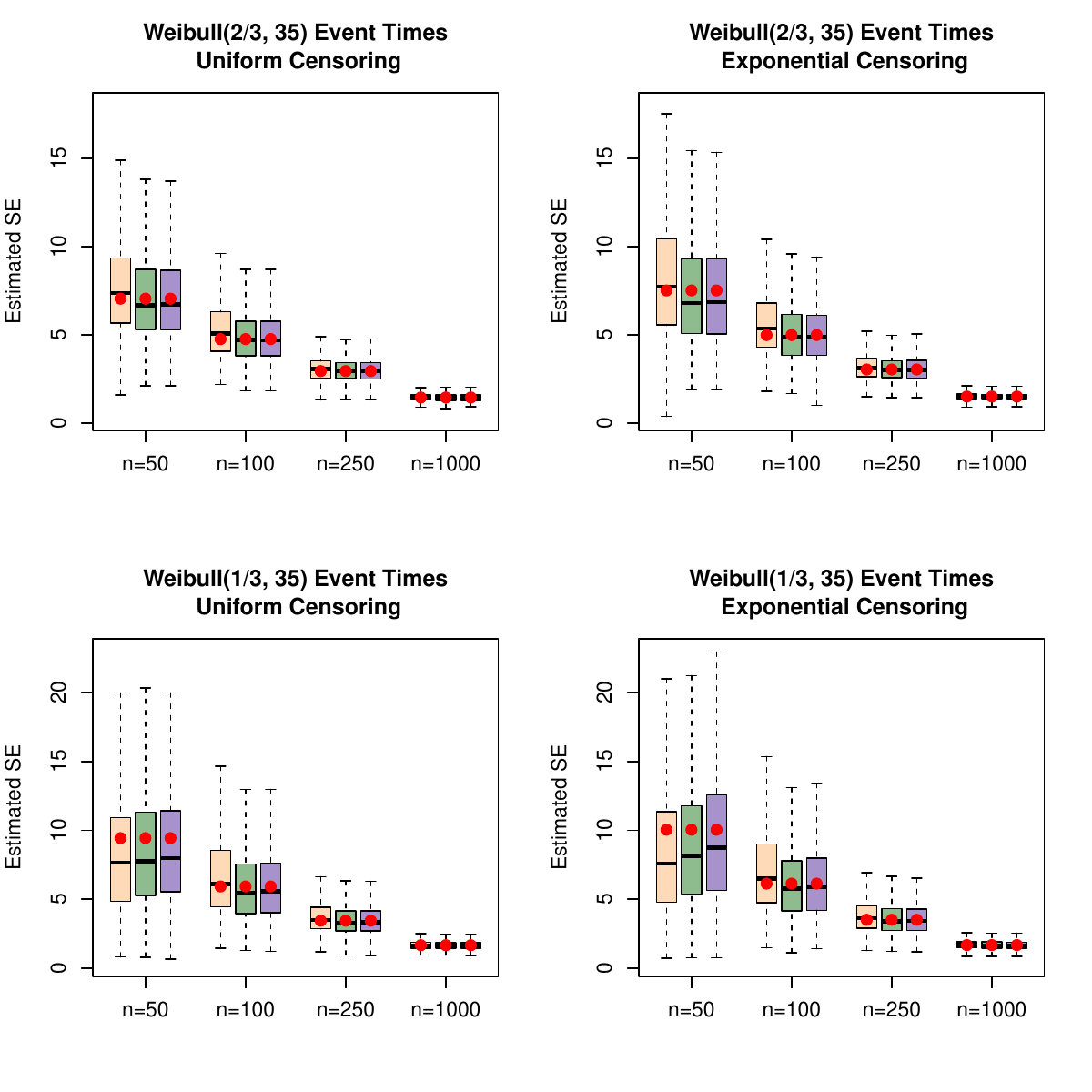}
    \caption{Study-level simulation results for the scenarios with the highly skewed Weibull event time distribution. The box plots illustrate the estimated standard errors (SEs) of the median survival time from the Wald approximation-based approach. The peach boxes correspond to when the Brookmeyer-Crowley method based on a log transformation was used to construct the 95\% confidence interval; The green boxes correspond to the Brookmeyer-Crowley method based on a log-minus-log transformation; The purple boxes correspond to the nonparametric bootstrap method. The true standard errors are illustrated by red dots.  \label{fig:sim res study-level skew}}
\end{figure}

The simulation results are presented in Figure \ref{fig:sim res study-level skew} and Table \ref{tab: sim res study-level skew}. The relative bias of the standard error estimators in scenarios with the $\mathrm{Weibull}(k = 2/3, \lambda=35)$ event time distribution were comparable to that observed in the simulations in the main text. However, for the more highly skewed $\mathrm{Weibull}(k = 1/3, \lambda=35)$ event time distribution, we observed larger relative bias values especially when the Brookmeyer-Crowley method with the log transformation was used to construct confidence intervals. For example, in the scenario with exponential censoring and a sample size of 100, the relative bias reached -19\%. Consistent with the simulation results from the main text, the relative bias decreased as the sample size increased. When the sample size was 1000, the relative bias was below 2\% across all scenarios.

\begin{table}[H]
\caption{Relative bias for estimating the standard error of median survival for the scenarios with the highly skewed Weibull event time distribution. The ``CI Method" column describes the method used to construct the 95\% confidence interval around median survival. In this column, ``BC (log)" denotes the Brookmeyer-Crowley method with the log transformation and ``BC (log-log)" denotes the Brookmeyer-Crowley method with the log-minus-log transformation.} \label{tab: sim res study-level skew}
\begin{center}
\begin{tabular}{lllllll}
\hline
& & & \multicolumn{4}{c}{Sample Size} \\ \cline{4-7}
Event Time Dist. & Censoring Dist. & CI Method & $n = 50$ & $n = 100$ & $n = 250$ & $n = 1000$ \\ 
  \hline
$\mathrm{Weibull}(2/3,35)$ & Uniform & BC (log) & 10 & 12 & 5 & 1 \\ 
 & & BC (log-log) & 3 & 3 & 2 & 0 \\ 
 & & Bootstrap & 3 & 3 & 2 & 0 \\ 
 & Exponential & BC (log) & 12 & 16 & 6 & 0 \\ 
 & & BC (log-log) & 2 & 4 & 2 & -1 \\ 
 & & Bootstrap & 2 & 3 & 2 & -1 \\ 
$\mathrm{Weibull}(1/3,35)$ & Uniform & BC (log) & -12 & 16 & 9 & 1 \\ 
 &  & BC (log-log) & -7 & 5 & 3 & 0 \\ 
 &  & Bootstrap & -7 & 6 & 4 & 0 \\ 
 & Exponential & BC (log) & -15 & 19 & 9 & 2 \\ 
  &  & BC (log-log) & -8 & 5 & 3 & 1 \\ 
   &  & Bootstrap & -4 & 8 & 3 & 0 \\ 
   \hline
\end{tabular}
\end{center}
\end{table}

\subsection{Impact of the number of bootstrap replicates}

We adopt the simulation setup as in Section 3.1.1 of the main text. For computational ease, we fixed the event time distribution to be the exponential distribution and fixed the censoring distribution to be the uniform distribution. We considered that studies report percentile-based nonparametric bootstrap confidence intervals with 1000, 2000, 5000, and 10000 bootstrap replicates. 

\begin{figure} [H]
    \centering
\includegraphics[width=0.6\textwidth]{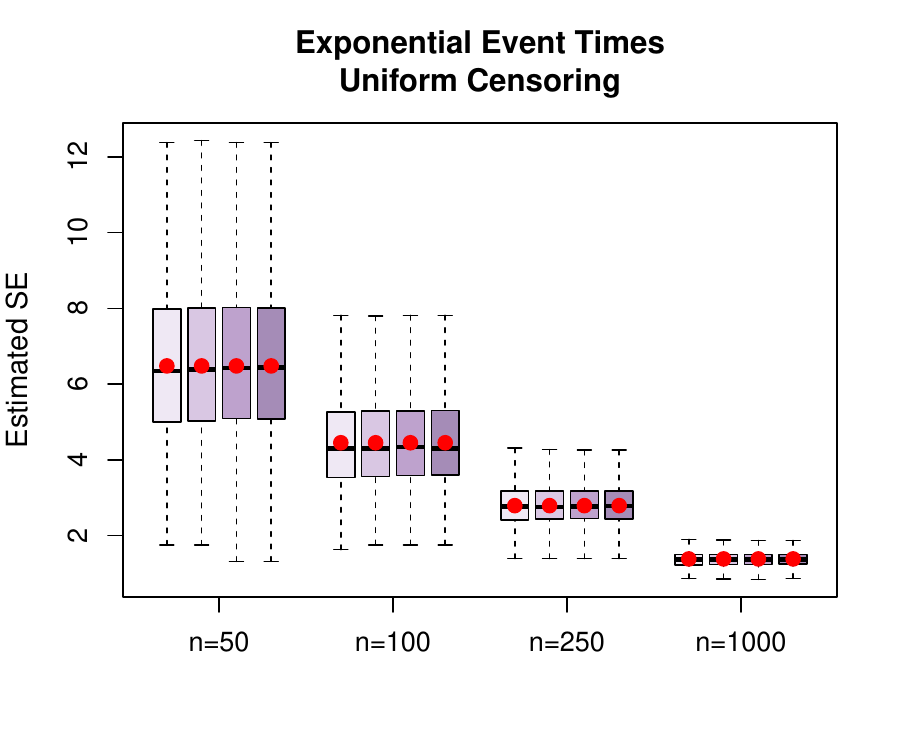}
    \caption{Study-level simulation results when varying the number of bootstrap replicates. The box plots illustrate the estimated standard errors (SEs) of the median survival time from the Wald approximation-based approach when the nonparametric bootstrap method was used. The lightest shade boxes correspond to 1000 bootstrap replicates. The progressively darker shaded boxes correspond to 2000, 5000, and 10000 bootstrap replicates. The true standard errors are illustrated by red dots.  \label{fig:sim res study-level bootstrap}}
\end{figure}

We applied the Wald approximation to estimate the standard error of the Kaplan-Meier median survival estimator based on each of these four confidence intervals. The results are summarized in Figure \ref{fig:sim res study-level bootstrap}. The standard error estimates were not strongly affected by the number of bootstrap replicates. Tables \ref{tab: sim res study-level bootstrap} and \ref{tab: sim res study-level bootstrap se} present the relative bias and standard error of the Wald approximation, respectively.

\begin{table}[H]
\caption{Relative bias for estimating the standard error of median survival when varying the number of bootstrap replicates.} \label{tab: sim res study-level bootstrap}
\begin{center}
\begin{tabular}{lllll}
\hline
& \multicolumn{4}{c}{Sample Size} \\ \cline{2-5}
Number of Replicates & $n = 50$ & $n = 100$ & $n = 250$ & $n = 1000$ \\ 
  \hline
1000 & 3.7 & 1.0 & 0.9 & -1.2 \\ 
2000 & 3.9 & 1.4 & 0.9 & -1.1 \\ 
5000 & 4.3 & 1.6 & 1.4 & -0.8 \\ 
10000 & 4.1 & 1.7 & 1.3 & -0.7 \\ 
   \hline
\end{tabular}
\end{center}
\end{table}

\begin{table}[H]
\caption{Standard error for estimating the standard error of median survival when varying the number of bootstrap replicates.} \label{tab: sim res study-level bootstrap se}
\begin{center}
\begin{tabular}{lllll}
\hline
& \multicolumn{4}{c}{Sample Size} \\ \cline{2-5}
Number of Replicates & $n = 50$ & $n = 100$ & $n = 250$ & $n = 1000$ \\ 
  \hline
1000 & 2.45 & 1.30 & 0.57 & 0.20 \\ 
2000 & 2.42 & 1.30 & 0.57 & 0.20 \\ 
5000 & 2.41 & 1.31 & 0.57 & 0.20 \\ 
10000 & 2.40 & 1.32 & 0.57 & 0.20 \\ 
   \hline
\end{tabular}
\end{center}
\end{table}

\subsection{Obtaining true standard errors}

In this subsection, we present additional details on the approach used to obtain the true standard error values used in our simulations. 

We performed Monte Carlo integration with $10^5$ samples \cite{naimi2025computing}. This approach involves iteratively performing the following steps $10^5$ times: For iteration $i$,
\begin{enumerate}
    \item Simulate a dataset under the true data generating mechanism. 
    \item Compute the Kaplan-Meier estimate of median survival based on the dataset in Step 1, denoted by $\hat{\theta}_i$.
\end{enumerate}
The standard deviation of the $\hat{\theta}_i$ values is taken as the true value of the standard error of the Kaplan-Meier estimator.

\section{Additional materials for the meta-analytic level simulations}

\subsection{Misspecifying the form of between-study heterogeneity}  \label{appendix: simulations meta-analytic level}

In this section, we conduct simulations that evaluate the performance of the Wald approximation-based approach when misspecifying the form of the between-study heterogeneity. One motivation for these simulations arises from the fact that the method for meta-analyzing the difference of medians involves different distributional assumptions on the between-study heterogeneity compared to the method for meta-analyzing the ratio of medians, yet researchers may wish to apply both methods to the same data (e.g., see Section 4 in the main text). These simulations involve meta-analyzing the \emph{difference of medians} in scenarios where the heterogeneity assumptions for meta-analyzing the \emph{ratio of medians} (rather than the difference of medians) actually hold. Additionally, these simulations involve meta-analyzing the \emph{ratio of medians} in scenarios where the heterogeneity assumptions for meta-analyzing the \emph{difference of medians} hold. 

\subsubsection{Meta-analysis of the difference of median survival}

\begin{figure} [H]
    \centering
\includegraphics[width=0.5\textwidth]{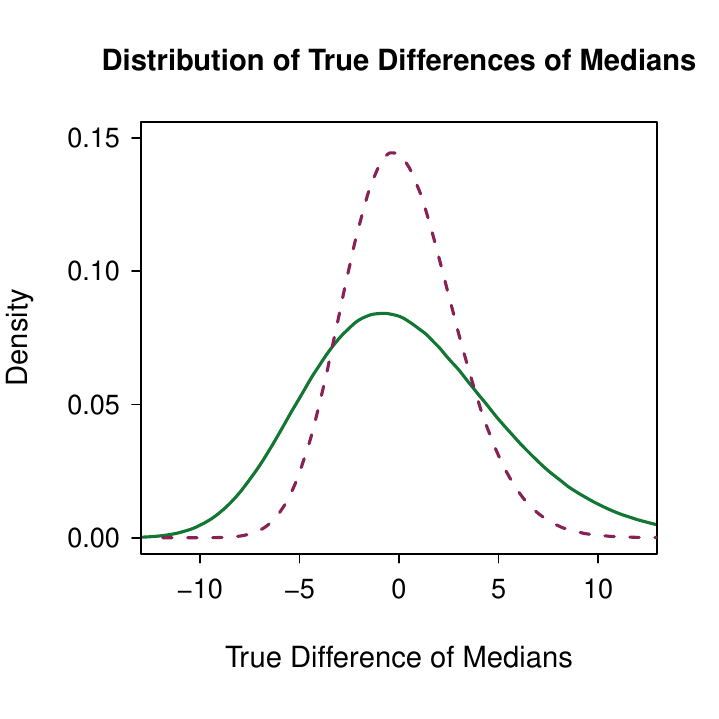}
    \caption{Probability density function of the distribution of the true study-specific differences of medians. The dark green lines correspond to the settings with high between-study heterogeneity, and the dark purple lines correspond to the settings with low between-study heterogeneity. \label{fig:distribution true effects dom}}
\end{figure}

\paragraph{Data generation} 
We used the data generating mechanism described in Section 3.2.3 of the main text. In these settings, recall that the true study-specific log ratios of medians are normally distributed across studies. The distributions of the true study-specific differences of medians in these settings are illustrated in Figure \ref{fig:distribution true effects dom}, which were obtained by Monte Carlo integration with $10^7$ samples. 

Note that the degree of between-study variance was higher in these simulations compared to those in Section 3.2.2 of the main text. The variance of the true study-specific difference of medians was approximately 24.14 in the high heterogeneity setting and 7.80 in the low heterogeneity setting here. For comparison, recall that the variance of the true study-specific differences of medians was 12 in the high heterogeneity setting and 4 in the low heterogeneity setting in the simulations in Section 3.2.2.

We applied the Wald approximation-based approach to meta-analyze the difference of medians. We also applied the benchmark approach which uses the true standard errors of the study-specific differences of medians, as described in the main text. Like the Wald approximation-based approach, note that the benchmark approach also misspecifies the distribution study-specific difference of medians in these simulations. That is, the benchmark approach represents the best-case scenario of the Wald approximation-based approach with respect to estimating the within-study standard errors.

\begin{figure}[H] 
  \centering
   \includegraphics[width=\textwidth]{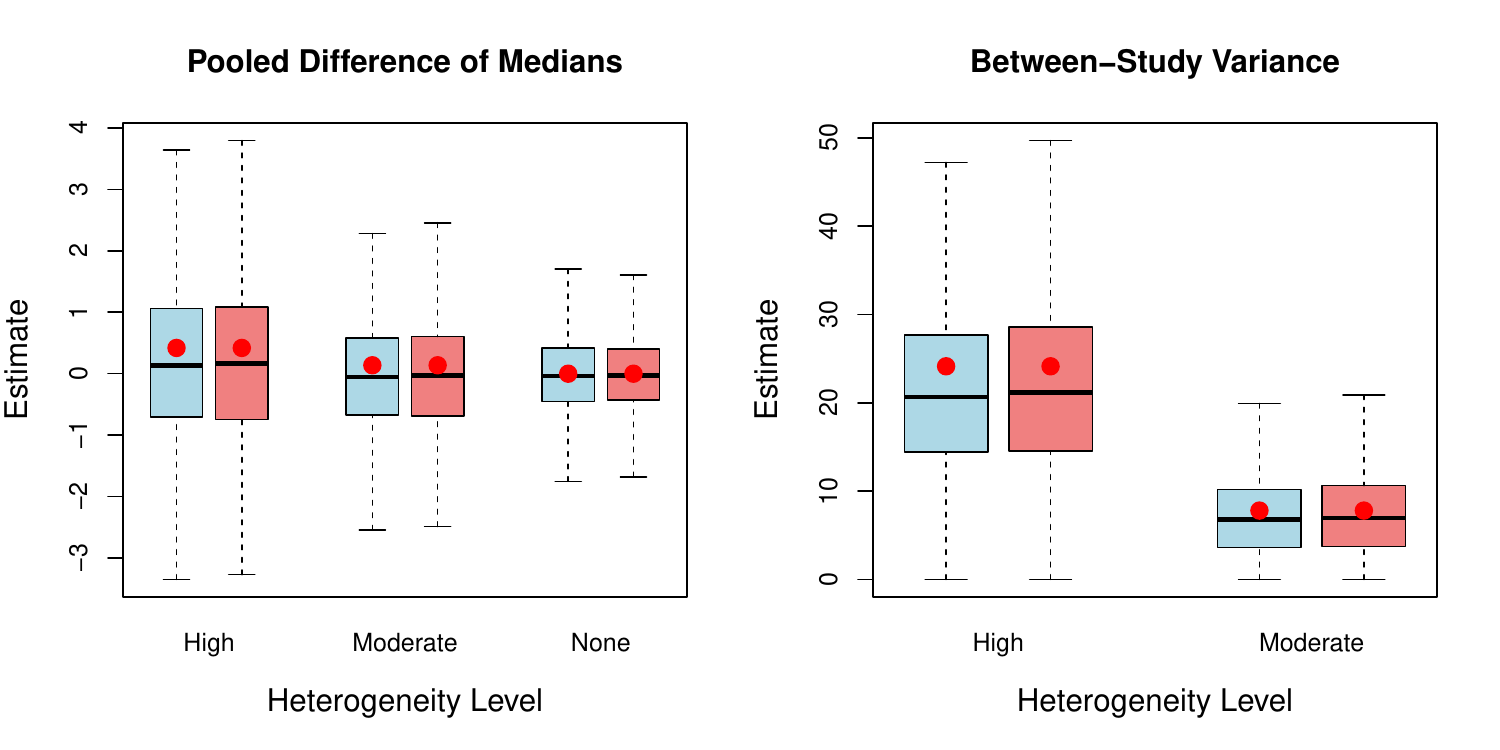}
   \caption{Estimates of the pooled difference of medians (left panel) and the between-study variance (right panel). The blue boxes correspond to the Wald approximation-based approach, and the red boxes correspond to the benchmark approach. The red dots indicate the true values.} \label{fig: sim res median difference wrongtransform}
\end{figure}

\paragraph{Results} 
The results are summarized in Figure \ref{fig: sim res median difference wrongtransform} and Table \ref{tab: sim res median difference wrongtransform}. The trends are similar to those of the simulation study in Section 3.2.2 of the main text. The Wald approximation-based approach performed similarly to the benchmark approach in each setting. As one would expect, the standard errors of the estimators were generally larger here compared to those in Section 3.2.2 due to the higher degree of between-study heterogeneity.

\begin{table}[H]
\caption{Simulation results for estimating the pooled difference of median. The bias, standard error (SE), and coverage of the 95\% confidence intervals are reported for the Wald approximation-based approach and benchmark approach.} \label{tab: sim res median difference wrongtransform}
\begin{center}
\begin{tabular}{@{\extracolsep{6pt}}llllllll@{}}
\hline
& & \multicolumn{3}{c}{Wald Approx.\ Approach} & \multicolumn{3}{c}{Benchmark Approach} \\ \cline{3-5} \cline{6-8}
Target Parameter & Heterogeneity & Bias & SE & Coverage & Bias & SE &  Coverage \\
  \hline
Pooled Dif.\ of Medians & High & -0.26 & 1.33 & 0.93 & -0.25 & 1.34 & 0.93 \\ 
& Moderate & -0.17 & 0.92 & 0.94 & -0.16 & 0.93 & 0.94 \\ 
& None & -0.02 & 0.63 & 0.96 & -0.02 & 0.63 & 0.96 \\    \addlinespace
Between-Study Variance & High & -2.10 & 10.73 & 0.95 & -1.51 & 10.98 & 0.95 \\ 
& Moderate & -0.43 & 5.08 & 0.96 & -0.18 & 5.22 & 0.95 \\ 
   \hline
\end{tabular}
\end{center}
\end{table}

\subsubsection{Meta-analysis of the ratio of median survival} 

\begin{figure} [H]
    \centering
\includegraphics[width=0.5\textwidth]{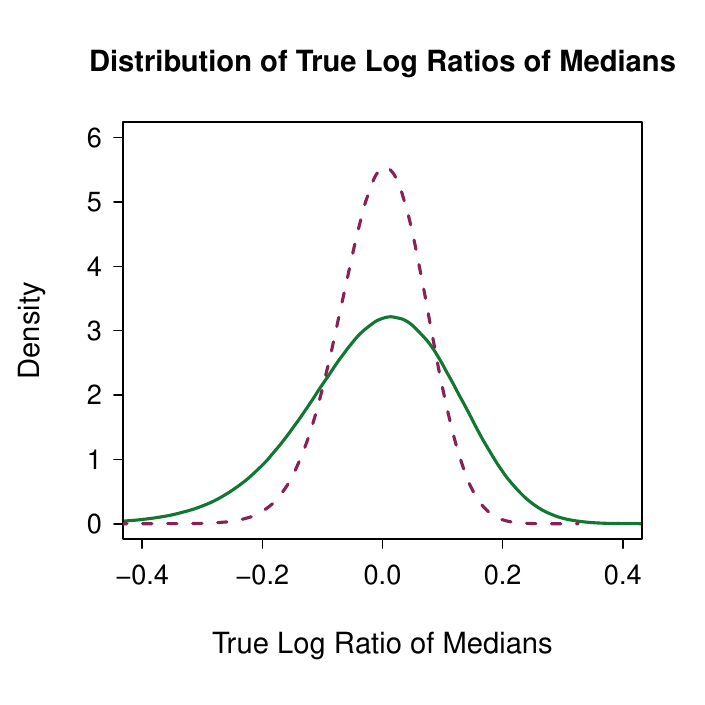}
    \caption{Probability density function of the distribution of the true study-specific log ratios of medians. The dark green lines correspond to the settings with high between-study heterogeneity, and the dark purple lines correspond to the settings with low between-study heterogeneity. \label{fig:distribution true effects rom}}
\end{figure}

\paragraph{Data generation} 
We used the data generating mechanism described in Section 3.2.2 of the main text. Here, the true study-specific differences of medians are normally distributed across studies. The distributions of the true study-specific log ratios of medians in these settings are illustrated in Figure \ref{fig:distribution true effects rom}, which were obtained by Monte Carlo integration with $10^7$ samples. 

The degree of between-study variance is lower in these simulations compared to those in Section 3.2.3 of the main text. The variance of the true study-specific log ratio of medians is approximately 0.016 in the high heterogeneity setting and 0.005 in the low heterogeneity setting here. Recall that the variance of the true study-specific log ratio of medians was 0.03 in the high heterogeneity setting and 0.01 in the low heterogeneity settings in the simulations in Section 3.2.3.

We applied the Wald approximation-based approach and benchmark approach to meta-analyze the ratio of medians.

\begin{figure}[H] 
  \centering
   \includegraphics[width=\textwidth]{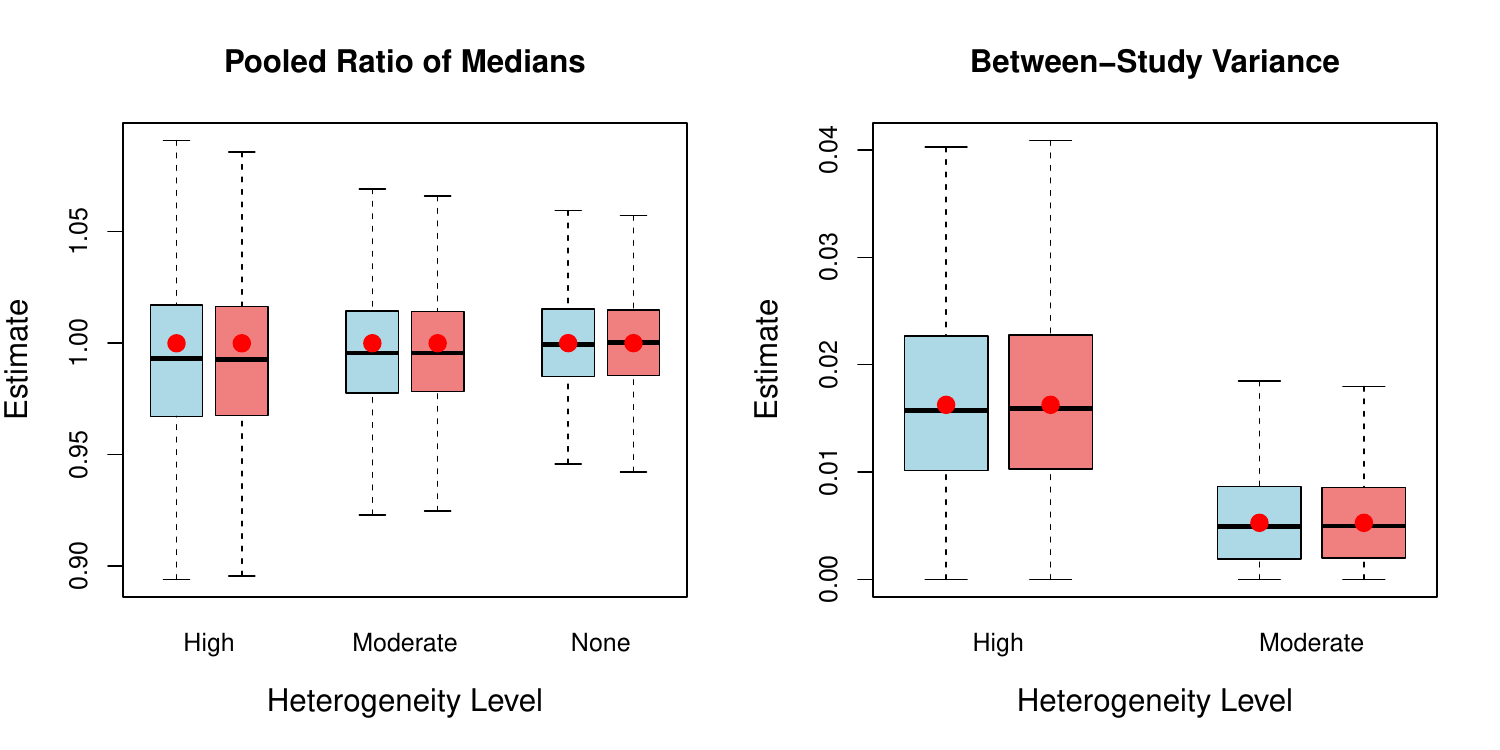}
   \caption{Estimates of the pooled ratio of medians (left panel) and the between-study variance (right panel). The blue boxes correspond to the Wald approximation-based approach, and the red boxes correspond to the benchmark approach. The red dots indicates the true values.} \label{fig: sim res median ratio wrongtransform}
\end{figure}

\paragraph{Results} 
The results are summarized in Figure \ref{fig: sim res median ratio wrongtransform} and Table \ref{tab: sim res median ratio wrongtransform}. We observed similar trends as those in Section 3.2.3 of the main text. The Wald approximation-based approach and benchmark approach performed similarly. As expected, the standard errors of the estimators were generally smaller here compared to those in Section 3.2.3 due to the lower degree of between-study heterogeneity. 

\begin{table}[H]
\caption{Simulation results for estimating the pooled ratio of medians. The bias, standard error (SE), and coverage of the 95\% confidence intervals are reported for the Wald approximation-based approach and benchmark approach. The bias and standard error values are multiplied by 100.} \label{tab: sim res median ratio wrongtransform}
\begin{center}
\begin{tabular}{@{\extracolsep{6pt}}llllllll@{}}
\hline
& & \multicolumn{3}{c}{Wald Approx.\ Approach} & \multicolumn{3}{c}{Benchmark Approach} \\ \cline{3-5} \cline{6-8}
Target Parameter & Heterogeneity & Bias & SE & Coverage & Bias & SE &  Coverage \\
  \hline
Pooled Ratio of Medians & High & -0.73 & 3.61 & 0.96 & -0.73 & 3.58 & 0.97 \\ 
& Moderate & -0.34 & 2.86 & 0.95 & -0.35 & 2.82 & 0.96 \\ 
& None & 0.01 & 2.20 & 0.97 & 0.02 & 2.16 & 0.97 \\    \addlinespace
Between-Study Variance & High & 0.07 & 0.98 & 0.94 & 0.08 & 0.96 & 0.94 \\ 
& Moderate & 0.05 & 0.48 & 0.96 & 0.05 & 0.48 & 0.96 \\ 
   \hline
\end{tabular}
\end{center}
\end{table}

\subsection{Impact of small within-study sample sizes}

\subsubsection{Meta-analysis of median survival}

The Wald approximation-based approach had poorer performance in these settings than those with larger sample sizes. The method was slightly biased for estimating the pooled median, resulting in coverage probabilities of 0.89, 0.90, and 0.65 for the pooled median in the scenarios with high, moderate, and no between-study heterogeneity, respectively. The benchmark approach showed some bias for estimating the pooled median, although its coverage was often nominal or near nominal.

The Wald approximation-based approach generally overestimated the between-study variance. For example, these methods had a bias of 12.49 and 7.50, respectively, in the scenario with high heterogeneity (i.e., $\tau^2 = 12$). The distributions of the between-study variance estimates were highly right-skewed, where some estimates exceeded 70. These methods consequently showed below nominal coverage for the between-study variance, with coverage probabilities ranging from 0.81 to 0.88. The Wald approximation-based approach generally performed worse for estimating between-study heterogeneity, likely due to underestimating the within-study variance. 

The poor coverage in the common effect scenario and overestimation of between-study variance may be reasonably expected for the Wald approximation-based approach. These can be seen as consequences of the Wald approximation systematically underestimating the within-study variances, as found in the study-level simulations with small sample sizes. 

\begin{figure}[H] 
  \centering
   \includegraphics[width=0.975\textwidth]{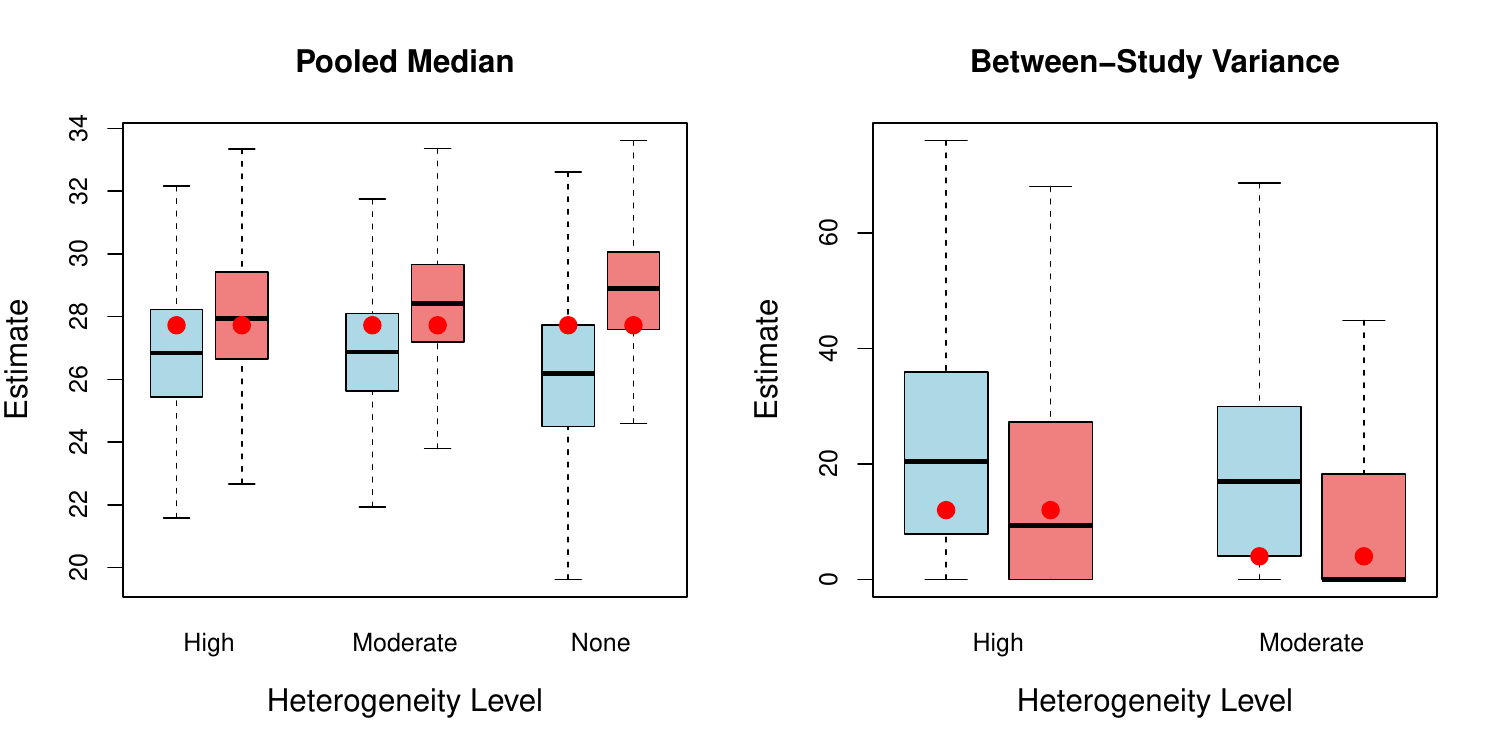}
   \caption{Estimates of the pooled median (left panel) and the between-study variance (right panel) in the simulation scenarios with small within-study sample sizes. The blue boxes correspond to the Wald approximation-based approach, and the red boxes correspond to the benchmark approach. The red dots indicate the true values} \label{fig: sim res median smalln}
\end{figure}

\begin{table}[H]
\caption{Simulation results for meta-analyzing median survival in the scenarios with small within-study sample sizes. The bias, standard error (SE), and coverage of the 95\% confidence intervals are reported for the Wald approximation-based approach and benchmark approach.} \label{tab: sim res median smalln}
\begin{center}
\begin{tabular}{@{\extracolsep{6pt}}llllllll@{}}
\hline
& & \multicolumn{3}{c}{Wald Approx.\ Approach} & \multicolumn{3}{c}{Benchmark Approach} \\ \cline{3-5} \cline{6-8}
Target Parameter &  Heterogeneity & Bias & SE & Coverage & Bias & SE &  Coverage \\
  \hline
Pooled Median & High & -0.85 & 2.06 & 0.89 & 0.35 & 2.08 & 0.96 \\ 
& Moderate & -0.83 & 1.91 & 0.90 & 0.78 & 1.90 & 0.97 \\ 
& None & -1.72 & 2.79 & 0.65 & 1.12 & 1.89 & 0.92 \\   \addlinespace
Between-Study Variance & High & 12.49 & 21.86 & 0.88 & 7.50 & 29.13 & 0.87 \\ 
& Moderate & 15.82 & 18.32 & 0.81 & 10.57 & 28.17 & 0.88 \\ 
   \hline
\end{tabular}
\end{center}
\end{table}

\subsubsection{Meta-analysis of the difference of median survival}

The Wald approximation-based approach and benchmark approach generally performed similarly to each other for estimating the pooled difference of medians. Both methods were approximately unbiased and the benchmark approach was slightly more efficient. While both methods often had nominal coverage of their 95\% confidence intervals for the pooled difference of medians, the Wald approximation-based approach had a coverage of 0.89 in the scenario with common effect meta-analyses.  scenario without between-study heterogeneity, which may be explained by this approach underestimating the within-study variances.

Both methods produced similar estimates of between-study variance. They both had some degree of bias and  overestimated between-study variance. Consequently, both methods had slightly below nominal coverage of their confidence intervals for the between-study variance, with coverage probabilities between 0.90 and 0.91. 

\begin{figure}[H] 
  \centering
   \includegraphics[width=0.975\textwidth]{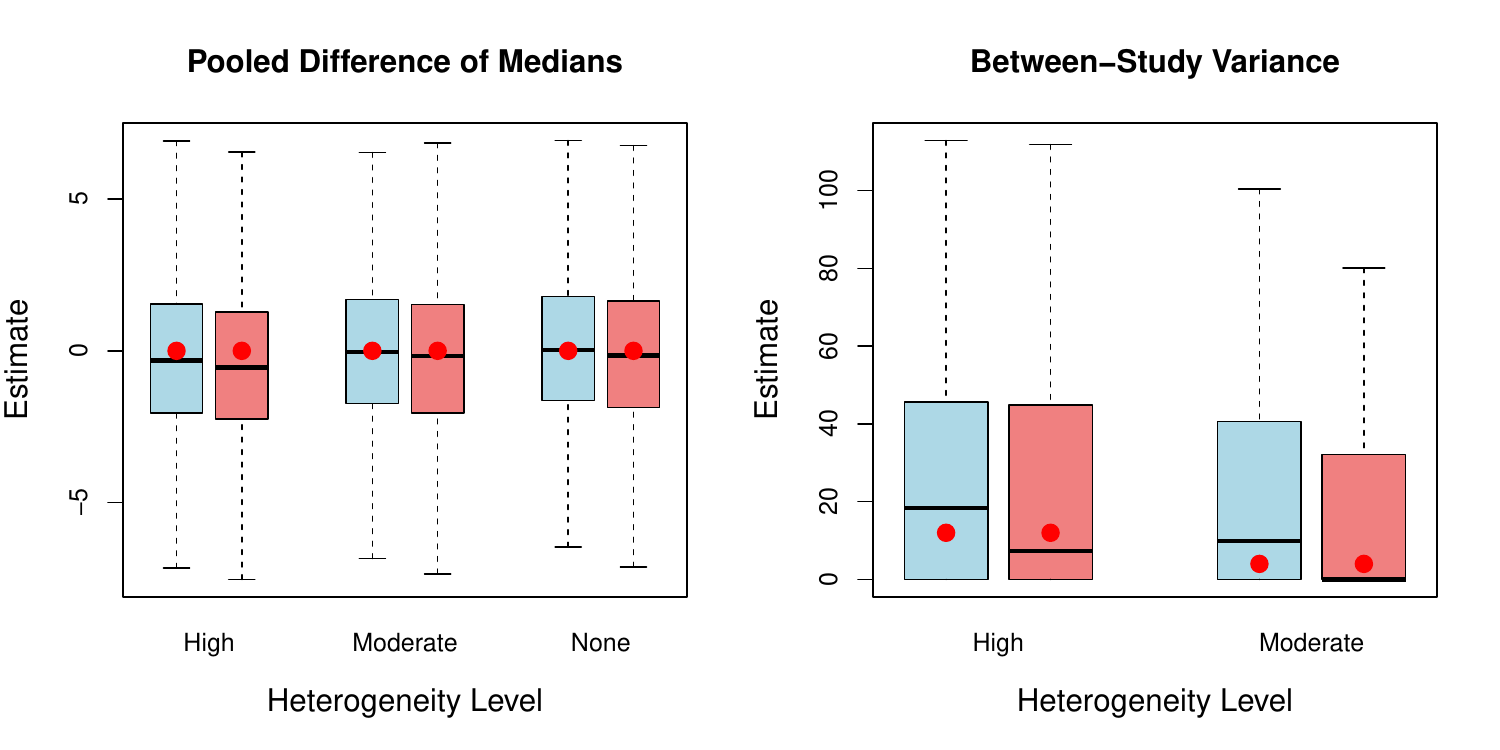}
   \caption{Estimates of the pooled difference of medians (left panel) and the between-study variance (right panel) in the simulation scenarios with small within-study sample sizes. The blue boxes correspond to the Wald approximation-based approach, and the red boxes correspond to the benchmark approach. The red dots indicate the true values} \label{fig: sim res median difference smalln}
\end{figure}

\begin{table}[H]
\caption{Simulation results for meta-analyzing the difference of median survival in the scenarios with small within-study sample sizes. The bias, standard error (SE), and coverage of the 95\% confidence intervals are reported for the Wald approximation-based approach and benchmark approach.} \label{tab: sim res median difference smalln}
\begin{center}
\begin{tabular}{@{\extracolsep{6pt}}llllllll@{}}
\hline
& & \multicolumn{3}{c}{Wald Approx.\ Approach} & \multicolumn{3}{c}{Benchmark Approach} \\ \cline{3-5} \cline{6-8}
Target Parameter & Heterogeneity & Bias & SE & Coverage & Bias & SE &  Coverage \\
  \hline
Pooled Dif.\ of Medians & High & -0.27 & 2.64 & 0.95 & -0.47 & 2.74 & 0.97 \\ 
& Moderate & -0.09 & 2.60 & 0.96 & -0.24 & 2.68 & 0.97 \\ 
& None & 0.05 & 2.82 & 0.89 & -0.09 & 2.57 & 0.96 \\    \addlinespace
Between-Study Variance & High & 18.00 & 36.42 & 0.91 & 17.58 & 44.61 & 0.91 \\  
& Moderate & 21.15 & 33.24 & 0.90 & 17.90 & 37.23 & 0.91 \\ 
   \hline
\end{tabular}
\end{center}
\end{table}

\subsubsection{Meta-analysis of the ratio of median survival}

Both methods produced similar estimates of the pooled ratio of medians. They were approximated unbiased and the benchmark approach had slightly greater efficiency. These methods often had nominal coverage of the their 95\% confidence intervals for the pooled ratio of medians. However, in the common effect meta-analysis, the Wald approximation-based approach had a coverage probability of 0.89, which may be explained by underestimating the within-study variances.  

The Wald approximation-based approach had more biased estimates of between-study heterogeneity compared to the benchmark approach. This resulted in below nominal coverage for the between-study variance for the Wald approximation-based approach.

\begin{figure}[H] 
  \centering
   \includegraphics[width=0.975\textwidth]{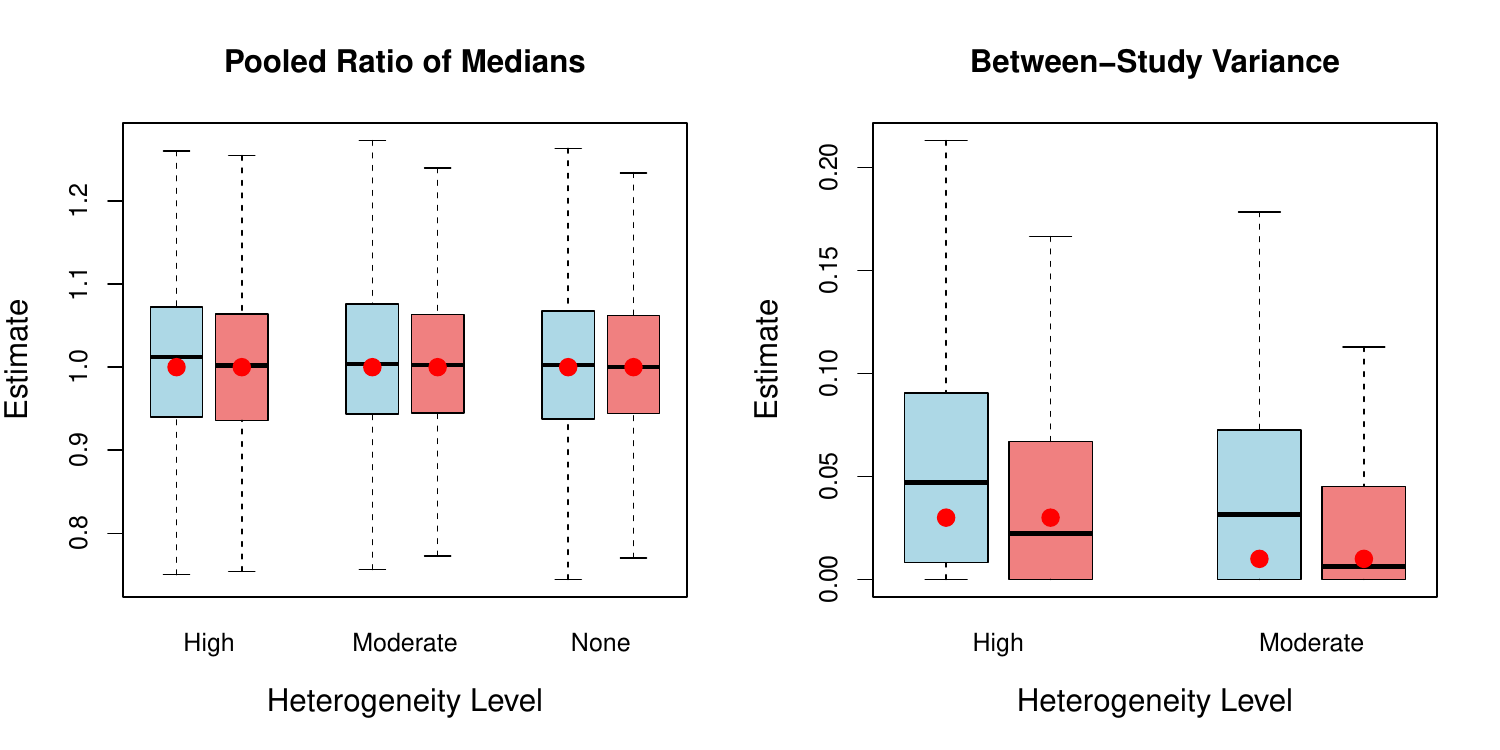}
   \caption{Estimates of the pooled ratio of medians (left panel) and the between-study variance (right panel) in the simulation scenarios with small within-study sample sizes. The blue boxes correspond to the Wald approximation-based approach, and the red boxes correspond to the benchmark approach. The red dots indicate the true values} \label{fig: sim res median ratio smalln}
\end{figure}

\begin{table}[H]
\caption{Simulation results for meta-analyzing the ratio of median survival in the scenarios with small within-study sample sizes. The bias, standard error (SE), and coverage of the 95\% confidence intervals are reported for the Wald approximation-based approach and benchmark approach. The bias and standard error values are multiplied by 100.} \label{tab: sim res median ratio smalln}
\begin{center}
\begin{tabular}{@{\extracolsep{6pt}}llllllll@{}}
\hline
& & \multicolumn{3}{c}{Wald Approx.\ Approach} & \multicolumn{3}{c}{Benchmark Approach} \\ \cline{3-5} \cline{6-8}
Target Parameter & Heterogeneity & Bias & SE & Coverage & Bias & SE &  Coverage \\
  \hline
Pooled Ratio of Medians & High & 0.87 & 9.84 & 0.96 & 0.25 & 9.31 & 0.96 \\ 
& Moderate & 0.71 & 9.61 & 0.96 & 0.42 & 8.87 & 0.97 \\ 
& None & 0.61 & 10.89 & 0.89 & 0.50 & 8.98 & 0.96 \\    \addlinespace
Between-Study Variance & High & 2.92 & 5.69 & 0.91 & 1.03 & 4.77 & 0.95 \\ 
& Moderate & 3.75 & 5.40 & 0.88 & 1.82 & 4.20 & 0.95 \\ 
   \hline
\end{tabular}
\end{center}
\end{table}

\section{Additional materials for the data application}\label{appendix: application}

\begin{table}[H]
\caption{Extracted summary data of overall survival (OS), in months, from the primary studies. The column titled ``Study" lists the National Clinical Trial (NCT) number. \label{tab: application data}}
\begin{center}
\begin{tabular}{@{\extracolsep{6pt}}llllll@{}} \hline
            & \multicolumn{2}{c}{Experimental Group} & \multicolumn{2}{c}{Comparator Group} \\ \cline{2-3} \cline{4-5}
            Study & $n$ & Median OS (95\% CI) & $n$ & Median OS (95\% CI) \\ \hline
NCT00946712 & 656 & 10.90 (9.50, 12.00) & 657 &  9.20 (8.70, 10.30) \\ 
  NCT01041781 & 154 & 11.40 (9.46, 14.06) & 158 & 12.50 (9.36, 13.83) \\ 
  NCT01386385 & 18 & 27.60 (17.40, 27.60) & 13 & 15.20 (6.60, 20.60) \\ 
  NCT01395758 & 51 &  6.80 (4.97, 10.70) & 45 &  8.50 (6.37, 13.97) \\ 
  NCT01466660 & 160 & 27.86 (25.13, 32.85) & 159 & 24.54 (20.57, 28.88) \\ 
  NCT01642251 & 64 & 10.30 (8.90, 12.00) & 64 &  8.90 (8.30, 11.30) \\ 
  NCT01828112 & 115 & 18.10 (13.40, 23.90) & 116 & 20.10 (11.90, 25.10) \\ 
  NCT01933932 & 254 &  8.70 (3.60, 16.80) & 256 &  7.90 (3.80, 20.10) \\ 
  NCT01951586 & 148 & 10.70 (8.54, 12.35) & 78 & 10.90 (9.26, 15.54) \\ 
  NCT02008227 & 425 & 13.80 (11.80, 15.70) & 425 &  9.60 (8.60, 11.20) \\ 
  NCT02041533 & 211 & 14.36 (11.60, 17.45) & 212 & 13.21 (10.68, 17.08) \\ 
  NCT02151981 & 279 & 26.80 (23.50, 31.50) & 140 & 22.50 (20.20, 28.80) \\ 
  NCT02152631 & 270 &  7.40 (6.50,  8.80) & 183 &  7.80 (6.40,  9.50) \\ 
  NCT02264990 & 298 & 12.10 (10.40, 14.90) & 297 & 12.10 (10.00, 13.70) \\ 
  NCT02296125 & 279 & 38.60 (34.50, 41.80) & 277 & 31.80 (26.60, 36.00) \\ 
  NCT02352948 & 62 & 11.70 (8.20, 17.40) & 64 &  6.80 (4.90, 10.20) \\ 
  NCT02352948 & 174 & 11.50 (8.70, 14.10) & 118 &  8.70 (6.50, 11.70) \\ 
  NCT02366143 & 359 & 19.20 (17.00, 23.80) & 337 & 14.70 (13.30, 16.90) \\ 
  NCT02367781 & 456 & 18.60 (15.80, 21.20) & 229 & 13.90 (12.00, 18.70) \\ 
  NCT02367794 & 343 & 14.20 (12.30, 16.80) & 340 & 13.50 (12.20, 15.10) \\ 
  NCT02387216 & 71 &  7.70 (3.60, 10.40) & 38 &  8.40 (5.80, 14.70) \\ 
  NCT02395172 & 396 & 10.60 (9.20, 12.30) & 396 &  9.90 (8.10, 11.90) \\ 
  NCT02409342 & 277 & 18.90 (13.40, 23.00) & 277 & 14.70 (11.20, 16.50) \\ 
  NCT02450539 & 106 &  7.00 (5.00,  8.78) & 53 & 12.39 (7.13, 15.98) \\ 
  NCT02453282 & 163 & 11.90 (9.00, 17.70) & 162 & 12.90 (10.50, 15.00) \\ 
  NCT02453282 & 163 & 16.30 (12.20, 20.80) & 162 & 12.90 (10.50, 15.00) \\ 
  NCT02542293 & 410 & 10.90 (9.30, 12.60) & 413 & 12.10 (10.30, 13.50) \\ 
  NCT02657434 & 292 & 17.50 (13.20, 19.60) & 286 & 13.60 (11.00, 15.70) \\ 
  NCT02785952 & 125 & 10.00 (8.00, 14.40) & 127 & 11.00 (8.60, 13.70) \\ 
  NCT02855125 & 64 &  7.92 (6.28, 10.78) & 64 &  9.82 (7.66, 13.40) \\ 
            \hline 
\end{tabular}
\end{center}
\end{table}

\bibliographystyle{unsrt}
\bibliography{references}